\documentclass[11pt]{article}

\usepackage{enumitem}
\setlist[enumerate]{align=left}

\usepackage[utf8]{inputenc}
\usepackage[T1]{fontenc}

\usepackage[utf8]{inputenc} 
\usepackage[T1]{fontenc}    
\usepackage{hyperref}       
\usepackage{url}            
\usepackage{booktabs}       
\usepackage{amsfonts}       
\usepackage{nicefrac}       
\usepackage{microtype}      
\usepackage{xcolor}
\usepackage{lipsum}
\usepackage[labelfont=bf]{caption}
\usepackage[subrefformat=parens]{subcaption}

\usepackage{graphicx}
\usepackage{amsthm}

\usepackage{algpseudocode}

\RequirePackage{amsmath,amsfonts,amssymb}

\newcommand{\etal}{\textit{et al.}}
\newcommand{\ie}{\textit{i.e., }}
\newcommand{\eg}{\textit{e.g., }}
\newcommand{\wrt}{\textit{w.r.t. }}

\usepackage{scicite}
\usepackage{cite}
\renewcommand*{\citedash}{\hbox{-}\penalty\citepunctpenalty}

\usepackage{times}
\usepackage{cleveref}
\crefname{equation}{}{}
\crefname{figure}{}{}

\usepackage[letterpaper,left = 0.85in, right = 0.8in, top = 0.8in,bottom = 00.8in ]{geometry}

\newenvironment{sciabstract}{%
\begin{quote} \bf}
{\end{quote}}

\title{PARC: Physics-Aware Recurrent Convolutional Neural Networks to Assimilate Meso-scale Reactive Mechanics of Energetic Materials}

\author
{Phong C. H. Nguyen,$^{1}$ Yen-Thi Nguyen,$^{2}$ Joseph B. Choi$^{1}$, \\
Pradeep K. Seshadri,$^{2}$ H.S. Udaykumar,$^{2,\ast}$ Stephen S. Baek$^{1,3,\ast}$ \\
\\
\normalsize{$^{1}$School of Data Science, University of Virginia }\\
\normalsize{Charlottesville, VA 22903, United States}\\
\normalsize{$^{2}$Department of Mechanical Engineering, University of Iowa }\\
\normalsize{Iowa City, IA 52242, United States}\\
\normalsize{$^{3}$Department of Mechanical and Aerospace Engineering, University of Virginia,}\\
\normalsize{Charlottesville, VA 22903, United States}\\
\\
\normalsize{$^\ast$Corresponding authors: hs-kumar@uiowa.edu, baek@virginia.edu}
}

\begin{document} 
\maketitle 

\begin{sciabstract}
   The thermo-mechanical response of shock-initiated energetic materials (EM) is highly influenced by their microstructures, presenting an opportunity to engineer EM microstructure in a "materials-by-design" framework. However, the current design practice is limited, as a large ensemble of simulations is required to construct the complex EM structure-property-performance linkages. We present the Physics-Aware Recurrent Convolutional (PARC) Neural Network, a deep-learning algorithm capable of learning the mesoscale thermo-mechanics of EM from a modest number of high-resolution direct numerical simulations (DNS). Validation results demonstrated that PARC could predict the themo-mechanical response of shocked EM with a comparable accuracy to DNS but with notably less computation time. The physics awareness of PARC enhances its modeling capabilities and generalizability, especially when challenged in unseen prediction scenarios. We also demonstrate that visualizing the artificial neurons at PARC can shed light on important aspects of EM thermos-mechanics and provide an additional lens for conceptualizing EM.
\end{sciabstract}



\baselineskip14pt


\flushbottom
\section*{Introduction}
Energetic materials (EM) such as propellants, explosives, and pyrotechnics are key components in many military and civilian applications. EMs are composites of organic crystals, plasticizers, metals, and other inclusions, forming complex microstructural morphologies, which strongly influence the properties and performance characteristics of these materials \cite{Baer2002}. For instance, the sensitivity to impact and shock loading—one of the key performance parameters for the design of safe and reliable EMs—is strongly influenced by their microstructures \cite{Perry2018,Mang2021,Mi2020}. Voids, cracks, and interfaces in EM microstructures are potential sites for energy localization, \ie the formation of high-temperature regions called ``hotspots'' \cite{Johnson1985,Frey1982,Massoni1999,Menikoff2004}. Such hotspots are considered to be critical if they grow and produce steady deflagration fronts \cite{Tarver1996}. If a sufficient number of such critical hotspots are generated in the microstructure, chemical energy release can be rapid enough to couple with the incident shock wave, initiating a detonation. Therefore, microstructural features localize energy release at hotspots and shock-microstructure interactions can lead to a shock-to-detonation transition in EMs.

Many mechanisms for hotspot formation have been identified \cite{Field1992}, including void collapse, plastic dissipation, intergranular or interface frictions, etc. Among these, under strong shock conditions, void collapse \cite{Menikoff2004} and shear-induced localization \cite{Frey1982} are predominating mechanisms for hotspot formation and therefore, will be the focus of this paper. Experiments \cite{Johnson1985} and simulations \cite{Massoni1999} have shown that the collapse of microscale voids in an EM can generate strong hotspots that can grow to combust the surrounding materials. Hotspots that are large enough and reach sufficiently high temperatures can initiate chemical reactions to release energy, strengthening the traveling shock and triggering a self-sustaining detonation \cite{Field1992}. Since there is a strong correlation between the microstructural morphologies and initiation sensitivity, in principle, one can engineer the sensitivity of EMs by manipulating their microstructures \cite{Chun2020}. This insight has inspired an active area of research to discover quantitative relationships between EM microstructures and their thermomechanical properties and performance. The discovered relationships are then used to engineer EMs with targeted sensitivity and energy delivery characteristics in a `materials-by-design' paradigm \cite{Bruck2007,Alberi2018,Niu2021}. 

The design loop for such a materials-by-design framework can be accelerated by employing predictive multi-scale simulations to obtain structure-property-performance (SPP) linkages for general EMs. Highly-resolved molecular \cite{Holian,Duarte,Das2021a} or continuum simulations \cite{Rai2015,Rai2017b,Rai2018} are usually necessary for accurate modeling of the energy localization phenomena in complex microstructures. However, such direct numerical simulations (DNS) can be computationally intensive, and extracting useful quantities of interest to inform the design may present challenges \cite{Perry2018,Sen2018a}. In this work, we propose a deep learning model called physics-aware recurrent convolutional (PARC) neural networks to predict the microscale reactive mechanics of shock-initiated EMs with quantitative fidelity comparable to the DNS of shocked microstructures but at a dramatically reduced computational cost.

Establishing SPP linkages is a central component in the materials-by-design framework for EMs. In the current practice, properties and performance of microstructures are determined experimentally \cite{Cummock2021a,Liu2019}, or calculated using multiscale numerical experiments \cite{Sen2018a}. In the latter approach, a large number of mesoscale simulations are conducted using imaged or synthetic micro\textit{structures}\cite{Chun2020,Roy2022,Nguyen2022GAN} to develop surrogate models \cite{Nassar2019,Roy2020a} for the sub-grid energy localization rate (\textit{property}), which is then used to produce macroscale predictions of sensitivity (\textit{performance}) by bridging the micro- and macro- length scales \cite{Sen2018a}. In previous works, surrogate models for mesoscale energy localization were constructed using the Gaussian Random Process techniques \cite{Sen2017}. Training data were obtained from ensembles of simulations performed with canonical void shapes (circular/elliptical) and other simplifying assumptions to reduce the model complexity and computational efforts. In this vein, Nassar \etal \cite{Nassar2019} developed a machine learning (ML) based surrogate model in which hotspot ignition and growth rates were expressed as a function of void size and applied shock loading. Roy \etal \cite{Roy2020a} developed a more sophisticated surrogate model spanning a larger parameter space, including not only the void sizes, but also the aspect ratios of voids, their orientations, and volume fractions. Other approaches to quantifying hotspot dynamics have also relied on idealized synthetic microstructures with voids represented by circles \cite{Mi2020}, ellipses \cite{Roy2020a}, and rectangles \cite{Rai2015}. Surrogate models derived in this manner \cite{Sen2018b,Das2018} were used to close the macroscale system of equations modeling the shock-to-detonation transition to determine the critical energy for initiation \cite{James1996} and the run-to-detonation distances \cite{Sen2018a}. Using this multiscale framework, the relationships among microstructural parameters, mesoscale dynamics (\eg evolution of the hotspot temperature or hotspot area) and the macroscale quantities of interest (QoI) \cite{Walker1976,Lee1980} for EM sensitivity were determined.

The use of idealized, canonical microstructures \cite{Roy2020a} or void collapse calculations on isolated single voids \cite{Nassar2019} led to simplified approaches, allowing for the reduction in the dimensionality of the material design space. In addition, such approaches also facilitate the investigation of SPP linkages of EMs with a tractable computational effort. However, such idealized representations are unable to capture the full complexity of morphological structures (\eg large, elongated cracks, tortuous voids) that influence the energy localization phenomena at the mesoscale \cite{Nguyen2022}. Furthermore, simplified surrogates may not adequately model the interactions among microstructural features, such as void-void interactions \cite{Roy2020a}, contorted/branched voids, and cracks. In addition, most surrogate models estimate the mesoscale QoI through an effective scalar value obtained by averaging or homogenization techniques; therefore, such approaches cannot represent high-fidelity details and spatiotemporal variations of the hotspot dynamics. Hence, detailed local phenomena such as the initiation of hotspots or the combination of multiple hotspots cannot be properly modeled. 

An alternative to such simplified representations of microstructural dynamics is to assimilate SPP linkages directly from non-idealized, high-resolution DNS performed on real, imaged microstructures \cite{Rai2015}. However, due to the highly transient thermo-mechanics of shocked heterogeneous EMs, DNS with non-idealized microstructures are computationally intensive and require extremely fine grid resolutions \cite{Roy2020b}. For instance, a single, well-resolved mesoscale simulation of the reactive response of microstructures can take hours to days on high-performance computing facilities. Moreover, since there is a large parameter space of stochastic micromorphology that needs to be examined during the design process, a vast number of experiments and simulations are required, leading to formidable costs for analyzing, modeling, and designing new material microstructures. 

To overcome these limitations, in this work, we propose a deep learning-based approach called PARC. PARC is trained to assimilate the behavior of evolving temperature and pressure fields of shocked heterogeneous EMs in time, using a small dataset obtained from DNS of high-resolution, image-derived microstructures \cite{Roy2020b}. In this approach, no simplification of microstructural morphology, data reduction, or simplifying assumptions 
is required. Also, the proposed method does not include any spatial and temporal averaging to assimilate the mesoscale dynamics. To achieve such a capability, PARC is designed to model governing differential equations of the material state (temperature and pressure) using convolutional neural networks (CNN) in a recurrent formulation which are solved via data-driven integrations. Such a differentiator-integrator architecture makes PARC `physics-aware' and interpretable, as its mathematical formulation is reminiscent of typical DNS solvers. The architecture design also separates PARC and typical input-output mapping approaches of common ML-based methods in materials science. 

After training, PARC can assimilate the hotspot dynamics of complex microstructural morphologies in shocked EM and provide a high-fidelity time-evolving prediction of temperature and pressure with an accuracy comparable to DNS. Moreover, PARC achieves a steep reduction of computation time, providing predictions nearly three orders of magnitude faster than DNS. Finally, PARC can illuminate the mechanisms of hotspot formation and growth in shocked EM microstructures, and identify the morphological features contributing most to hotspot formations via neural network visualization techniques as demonstrated later. This PARC-enabled capacity to run high-throughput, high-fidelity simulations, when combined with synthetic microstructure generation methods \cite{Chun2020,Roy2022,Nguyen2022GAN}, can accelerate the EM characterization processes. In addition, these capabilities enable rapid explorations of the vast configuration space of EM microstructures and thereby facilitate the discovery of better functional structures.

\begin{figure}[tb!]
\centering
\includegraphics[width=\linewidth]{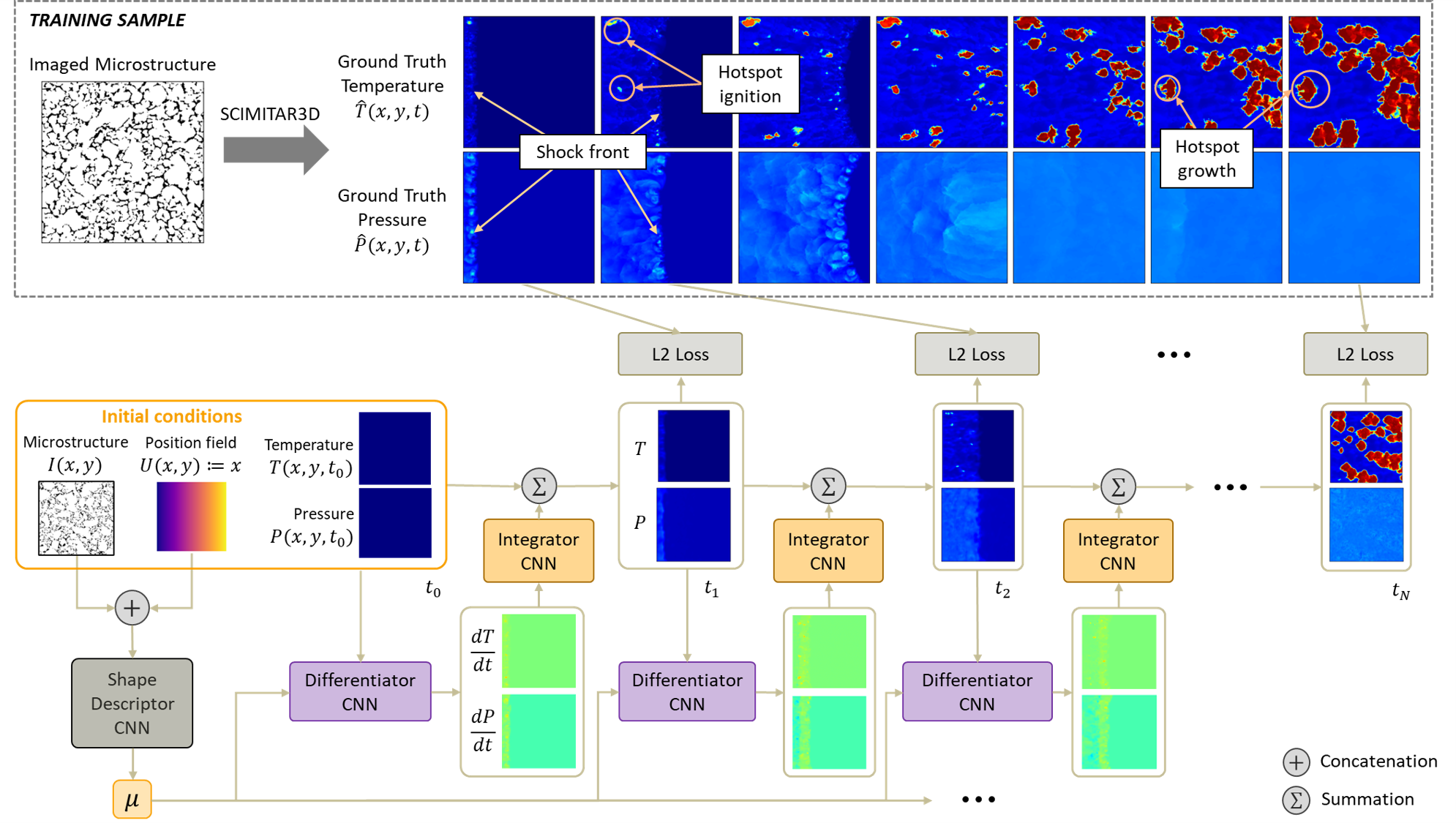}
\caption{\textbf{Overview of the PARC architecture.} Initially, the microstructure intensity field $I$ and the position field $U$ are encoded into the microstructure shape descriptor $\mu$ using the U-Net. The shape descriptor is sent to the differentiator CNN (purple box) along with the initial temperature and pressure fields to estimate the rate of change over time of the temperature and pressure at each grid point. The predicted temperature and pressure time derivatives are integrated over time using the integrator CNN (orange box) and added to the current temperature and pressure values to derive the QoI in the next time step. The process is repeated recursively and the networks parameter of differentiator and integrator are shared across all time steps. The temperature and pressure fields computed at each time step are combined to create time-evolving fields which assimilate the hotspot ignition and growth within shock-initiated EM microstructures.}
\label{fig:PARC-general}
\end{figure}

\section*{Results}
\subsection*{Physics-Aware Recurrent Convolutional Neural Network (PARC)}
Figure~\ref{fig:PARC-general} illustrates the overall architecture of PARC. The inputs to PARC include a binarized microstructure image (white pixels: materials; black pixels: voids), a position field, and initial conditions for temperature and pressure fields. The U-Net \cite{Ronneberger2015} subnetwork encodes the input microstructure image and position field into the microstructure shape descriptor $\mu$. This shape descriptor, along with the initial temperature and pressure fields, is the input to the \textit{differentiator} CNN (purple boxes in Fig.~\ref{fig:PARC-general}), which estimates the rates of change of the temperature and pressure field at each time step. The predicted rates of change are then integrated over time by the \textit{integrator} CNN (orange boxes in Fig.~\ref{fig:PARC-general}) to compute the temperature and pressure fields at the next time step. The process is repeated recursively until the simulation ends.

The state vector of the material system at a given position $\mathbf{r}=(x,y)$ and time t is denoted as $X(\mathbf{r};t):=[T(\mathbf{r};t),P(\mathbf{r};t)]^T$, where $T(\mathbf{r};t)$ and $P(\mathbf{r};t)$ are the temperature and pressure fields in the shocked EM microstructure sample. PARC models the evolution of the state vector, which contains the energy localization and the ignition and growth of hotspots generated by shock passage through an EM microstructure, via the following governing differential equation with initial conditions:

\begin{equation}
\label{eqn:prob_formulation}
    \left\{
    \begin{array}{ccl}
      \dfrac{d\mathbf{X}}{dt} &= &f(\mathbf{X}, \mu)    \\
      X\left(\mathbf{r};t=0\right) &= &\left[T_0\left(\mathbf{r}\right), P_0\left(\mathbf{r}\right)\right]    \\
    \end{array}
  \right.
\end{equation}

In Eq.~\ref{eqn:prob_formulation}, the shape descriptor, $\mu$, is introduced to account for the effect of microstructural morphology on the thermo-mechanical evolution of the state $\mathbf{X}$. Here, $t=0$ refers to the instant when a transient shock enters the microstructure, at which time the temperature field $T\left(\mathbf{r};t=0\right)$ and the pressure field $P\left(\mathbf{r};t=0\right)$ are assumed to be constant everywhere in the domain with the value of $T_0\left(\mathbf{r}\right) = 300 K$ and $P_0\left(\mathbf{r}\right) = 1 ATM$. Solving Eq.~\ref{eqn:prob_formulation} over a time interval $t\in[0,\tau]$ gives the temperature and pressure values $\mathbf{X}\left(\mathbf{r};\tau\right)$ at a time instance $t=\tau$:

\begin{equation}
\label{eqn:general_solution}
    \mathbf{X}(\mathbf{r};\tau)=\mathbf{X}_0 (\mathbf{r})+\int_{0}^{\tau}{f(\mathbf{X},\mu)dt}
\end{equation}
In Eq.~\ref{eqn:general_solution}, the computation of the integral $\int_{0}^{\tau}f\left(\mathbf{X},\mu \right)dt$ is prone to numerical integration errors, as the dynamics is violent and highly transient. There are also spatiotemporal discontinuities (material-void interfaces, shocks, reactive fronts) and large deformations (void collapse, material advection) inherent to the thermo-mechanics of shocked EMs. Therefore, during the shock-induced reaction of EMs, temperature and pressure fields exhibit high spatiotemporal gradients, resulting in temporal and spatial nonlinearities in the evolution described in Eq.~\ref{eqn:general_solution}, particularly for long simulation time. For large values of $\tau$, predictions from Eq.~\ref{eqn:general_solution} may diverge from the ground truth (taken to be the corresponding DNS fields). To circumvent this, the integral in Eq.~\ref{eqn:general_solution} is decomposed into discrete time intervals, in the form:

\begin{equation}
\label{eqn:discrete_sol}
    \arraycolsep=1.4pt\def\arraystretch{1.0}
    \begin{array}{ccl}
      \mathbf{X}\left(\mathbf{r};t_k\right) &= &\mathbf{X}_0\left(\mathbf{r}\right)+\int_{t_0}^{t_{k-1}}f\left(\mathbf{X},\mu\right)dt+\int_{t_{k-1}}^{t_k}f\left(\mathbf{X},\mu\right)dt    \\
      \\
       & = &\mathbf{X}\left(\mathbf{r};t_{k-1}\right)+\int_{t_{k-1}}^{t_k}f\left(\mathbf{X},\mu\right)dt.    \\

    \end{array}
\end{equation}
Here, to relieve the notational burden, we represent the integral over the interval $\left[t_{k-1},t_k\right]$ by the functional $S$:

\begin{equation}
    S\left(f\right):= \int_{t_{k-1}}^{t_k}f\left(\mathbf{X},\mu\right)dt.
\end{equation}
Note that both the time derivative $f\left(\mathbf{X},\mu\right)$ and the integral $\mathbf{S}\left(f\right)$ are mathematical operators, which can be modeled by CNNs as suggested by the universal approximation theorem~\cite{Lu2021}. Therefore, Eq.~\ref{eqn:discrete_sol} can be rewritten as:

\begin{equation}
\label{eqn:final_form}
    \mathbf{X}\left(\mathbf{r};t_k\right)=\mathbf{X}\left(\mathbf{r};t_{k-1}\right)+S\left(f\left(\mathbf{X},\mu\ \right|\ \theta\right)\ \left|\ \varphi\right),
\end{equation}
where $\theta$ and $\varphi$ are neural network parameters (weights and biases) for $f$ and $S$, respectively. Here, we refer to the neural network modeling $f$ as the \textit{differentiator} CNN (purple boxes in Figure~\ref{fig:PARC-general}) and the another modeling $S$ as the \textit{integrator} CNN (orange boxes in Figure~\ref{fig:PARC-general}). During the training of neural networks, the objective is to find the optimal parameters $\theta^\ast$ and $\varphi^\ast$ yielding the most accurate prediction of the evolution of the state $\mathbf{X}\left(\mathbf{r};t\right)$.

The PARC architecture (Figure~\ref{fig:PARC-general}) is the realization of the above neural network models in Eq.~\ref{eqn:final_form}. First, the morphology parameter $\mu$ is derived from the microstructure image $\mathbf{I}\left(x,\ y\right)$ and the position map $\mathbf{U}\left(x,y\right)$. Here the position map $\mathbf{U}:(x,y)\mapsto x$ is introduced to encode the relative position of each grid location with respect to the left boundary of the image domain where the shock enters. The inclusion of the position map is to account for the inherent translational equivariance \cite{Lenc2015} of CNNs, which makes them incapable of distinguishing the relative position of grid locations with respect to the shock front. The vertical coordinate $y$ is dropped as the shock travels horizontally in our setting. To encode the morphology parameter $\mu$, we employ the U-Net architecture \cite{Ronneberger2015}, informed by previous success in employing the U-Net as a morphology descriptor \cite{Baek2019}. The U-Net architecture takes the microstructure image $\mathbf{I}$ and the position map $\mathbf{U}$ and returns a morphology descriptor $\mu\in\mathbb{R}^{128}$ as an output at each grid location (see Supplemental Figure~\ref{fig:PARC_arch}A for more details).

At each time step $t_{k>0}$, the differentiator network $f\left(\mathbf{X},\mu\ \right|\ \theta)$ takes the morphology descriptor $\mu$ and the temperature and pressure fields $\mathbf{X}_k=\left[T_k,\ P_k\right]^T$ of the current time step as inputs and calculate the time derivatives of temperature and pressure fields $d\mathbf{X}_k/dt=[dT_k/dt,dP_k/dt]^T$. The differentiator CNN is comprised of two ResNet blocks \cite{He2016} and a $7\times7$ convolution layer for feature extraction, followed by two $1\times1$ convolution layers and a $3\times3$ convolution layer for enhancing the details (super-resolution) \cite{Dong2015} (see Supplemental Figure~\ref{fig:PARC_arch}B for more details). The outputs of the differentiator, \ie  the time derivatives of temperature and pressure fields, are then sent to the integrator network $\mathbf{S}(f\left|\ \varphi\right)$, which calculates the total variation of temperature and pressure at each grid location after a time interval $\Delta t$. The outputs of the integrator are then added to the current temperature and pressure fields $\mathbf{X}_k$ to derive the future state $\mathbf{X}_{k+1}$ at the next time step $t_{k+1}$. The integrator CNN has a similar architecture to the differentiator CNN (see Supplemental Figure~\ref{fig:PARC_arch}C). 

The above-described computation of the differentiator and the integrator is repetitively applied in a recursive fashion, in which the resultant temperature and pressure fields at each time step serve as the inputs for the next time step. The recurrent solution in Eq.~\ref{eqn:final_form} starts with $t_k=t_1$ and follows through until the final time step $t_{k}=t_P$ is reached. Through this recurrent computation, PARC returns the predictions of temperature and pressure fields $\mathbf{X}_1,\cdots,\mathbf{X}_P$ at discrete time instants $t_1,\ \cdots,\ t_P$, respectively. Note that the same neural network parameters $\theta$ and $\varphi$ of the differentiator and integrator are shared across time steps, making the PARC architecture recurrent. Such a recurrent convolutional architecture modeling of the differential equation $f$ and the integral $S$ resembles the way in which dynamic systems are simulated in typical DNS solvers for time-dependent problems and thus makes PARC \textit{physics-aware}. This physics-aware architecture makes PARC more interpretable compared to other black-box ML methods. That is, in a typical black-box ML model, the prediction mechanism is mostly input-output regression without any involved physics-awareness and such an approach leads to unexplainable and unreliable prediction results. This issue can be overcome by our proposed PARC model as demonstrated later. For more details on the PARC architecture design and implementation, please see Supplementary Materials.

Finally, the training of PARC is cast as an optimization problem in which the goal is to accurately model the spatiotemporal evolution of both temperature and pressure fields in shock-initiated reaction simulations of EM microstructures:
\begin{equation}
\label{eqn:loss_function}
    L\left(\theta,\varphi | \hat{X}\right) = \sum_{t_k}{} {|| \hat{X}(\mathbf{r};t) - X_{k-1} - S(f(X,\mu | \theta)|\phi)||}_2 + \sum_{t_k}{} {|| \hat{\dot{X}}(\mathbf{r};t) - f(X,\mu | \theta)||}_2
\end{equation}
Here, the quantities with the hat represent the ground truth data derived from DNS.

\subsection*{Prediction Performance}
\begin{figure}[tb]
     \centering
     \includegraphics[width=\textwidth]{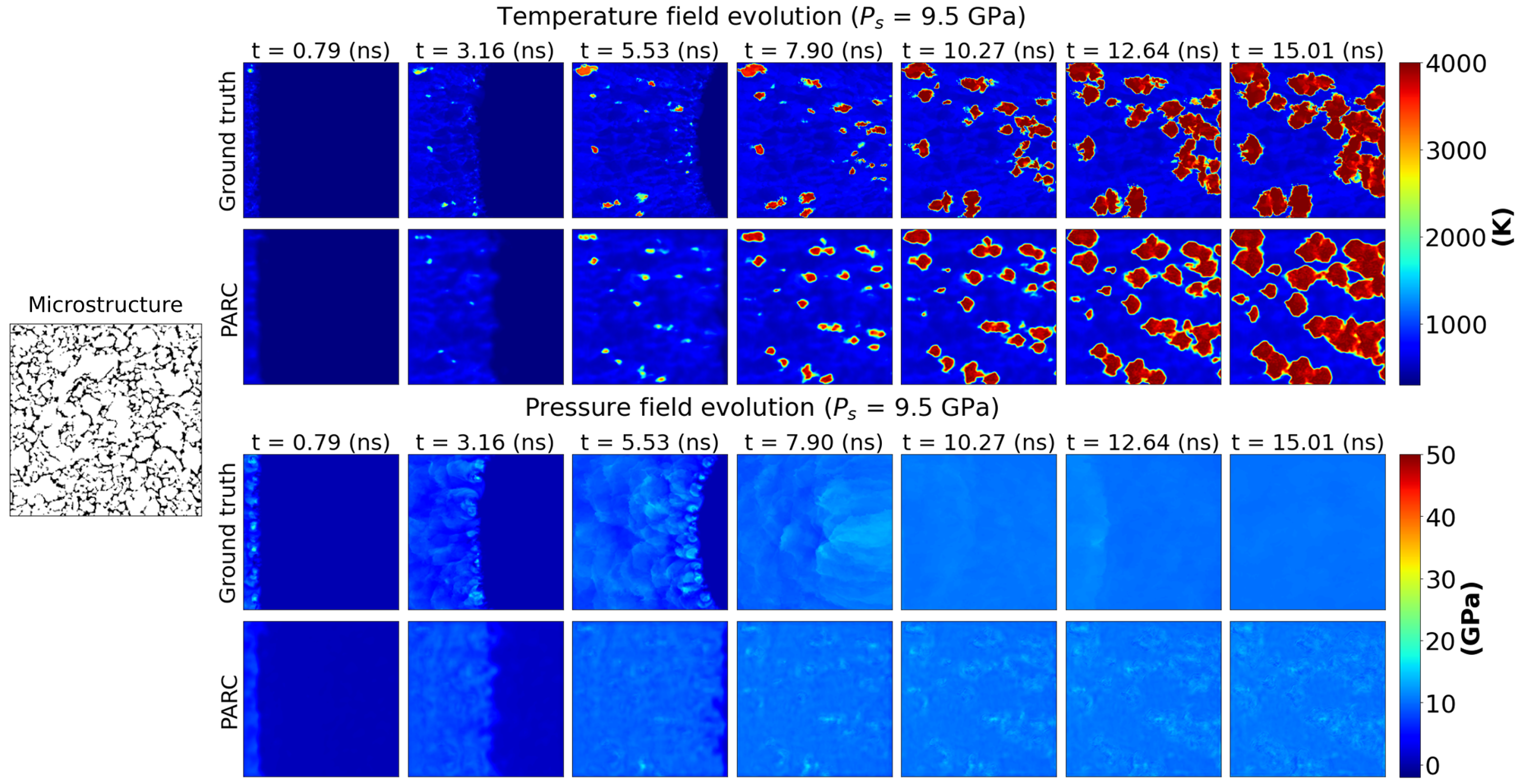}
\caption{\textbf{Temperature and pressure fields predicted by PARC displayed in comparison to DNS results.} The physics-aware architecture of PARC provides predictions of hotspot ignition and growth in good agreement with DNS.}
\label{fig:PARC-result}
\end{figure}

The ability of PARC to accurately predict the evolution of the temperature and pressure fields was evaluated in comparison with the corresponding DNS predictions. As depicted in Figure~\ref{fig:PARC-result} (and Supplementary Figures.~\cref{fig:test_1,fig:test_2,fig:test_3,fig:test_4,fig:test_5,fig:test_6,fig:test_7,fig:test_8} for more examples), PARC-predicted temperature and pressure fields were qualitatively comparable to those derived from the DNS. Notably, the locations, spatial patterns, and evolution of hotspots aligned well with those obtained from DNS. Similarly, the propagation of shock waves in the microstructure during the simulation was well captured, as indicated in both temperature and pressure field predictions.

For a more quantitative assessment, we evaluated the prediction accuracy of PARC via several EM sensitivity metrics that are known to be crucial for the design of EMs. Specifically, we measured the \textit{average hotspot temperature} $\overline{T}^{hs}$, the \textit{total hotspot area} ${A^{hs}}$, and their respective rates of change over time, $\dot{\overline{T}}^{hs}$ and $\dot{A}^{hs}$. The four QoIs are closely related to EM sensitivity. While the average hotspot temperature denotes the intensity of void collapse and the likelihood of the formation of ``critical'' hotspots, the total hotspot area shows the contribution of hotspots to energy localization in a control volume. Meanwhile, their rates of change over time are used to quantify the rate of energy deposition at the mesoscale. These four sensitivity QoIs are defined as:

\begin{equation}
\label{eqn:T_hs_bar}
    \overline{T}^{hs}\left(t_k\right)=\frac{\sum_{i=1}^{M}\sum_{j=1}^{N}\left(T_{ij}^{hs}\left(t_k\right)A_{ij}^{hs}\left(t_k\right)\right)}{A^{hs}\left(t_k\right)},
\end{equation}

\begin{equation}
\label{eqn:A_hs_bar}
    {A^{hs}}(t_k)=\sum_{i=1}^{M}\sum_{j=1}^{N}A_{ij}^{hs}(t_k),
\end{equation}

\begin{equation}
\label{eqn:T_hs_dot}
  \dot{\overline{T}}^{hs}(t_k)=\frac{\ \overline{T}^{hs}\left(t_k\right)-\overline{T}^{hs}\left(t_{k-1}\ \right)}{t_k-t_{k-1}},
\end{equation}

\begin{equation}
\label{eqn:A_hs_dot}
    \dot{A^{hs}}(t_k)=\frac{{A^{hs}}\left(t_j\right)-{A^{hs}}\left(t_{j-1}\right)}{t_k-t_{k-1}},
\end{equation}

\noindent where: 

\begin{equation}
\label{eqn:T_hs_compute}
T^{hs}_{ij}(t_k) = 
  \left\{
    \begin{array}{cl}
        T_{ij}(t_k) &\text{if}\quad T_{ij}(t_k)\geq875\ K,    \\
        0 & \text{otherwise,}   \\
     
    \end{array}
  \right.
\end{equation}

\begin{equation}
\label{eqn:A_hs_compute}
A^{hs}_{ij}(t_k)=
    \left\{
    \begin{array}{cl}
     A_{ij}(t_k) & \text{if}\quad T_{ij}(t_k)\geq875\ K,    \\
     0 & \text{otherwise.}   \\
    \end{array}
    \right.
\end{equation}

\noindent In Eqs.~\cref{eqn:T_hs_bar,eqn:A_hs_bar,eqn:T_hs_dot,eqn:A_hs_dot,eqn:T_hs_compute,eqn:A_hs_compute}, the subscript $ij$ indicates quantities specific to the grid location $(i,j)$ on a $M \times N$ grid. The superscript $hs$ indicates quantities specific to hotspots. $A_{ij}$ is the area of the grid cell occupied in the hotspot region, which is constant since a uniform grid was employed. 

\begin{figure}[t!]
    \centering
    \includegraphics[width=0.8\textwidth]{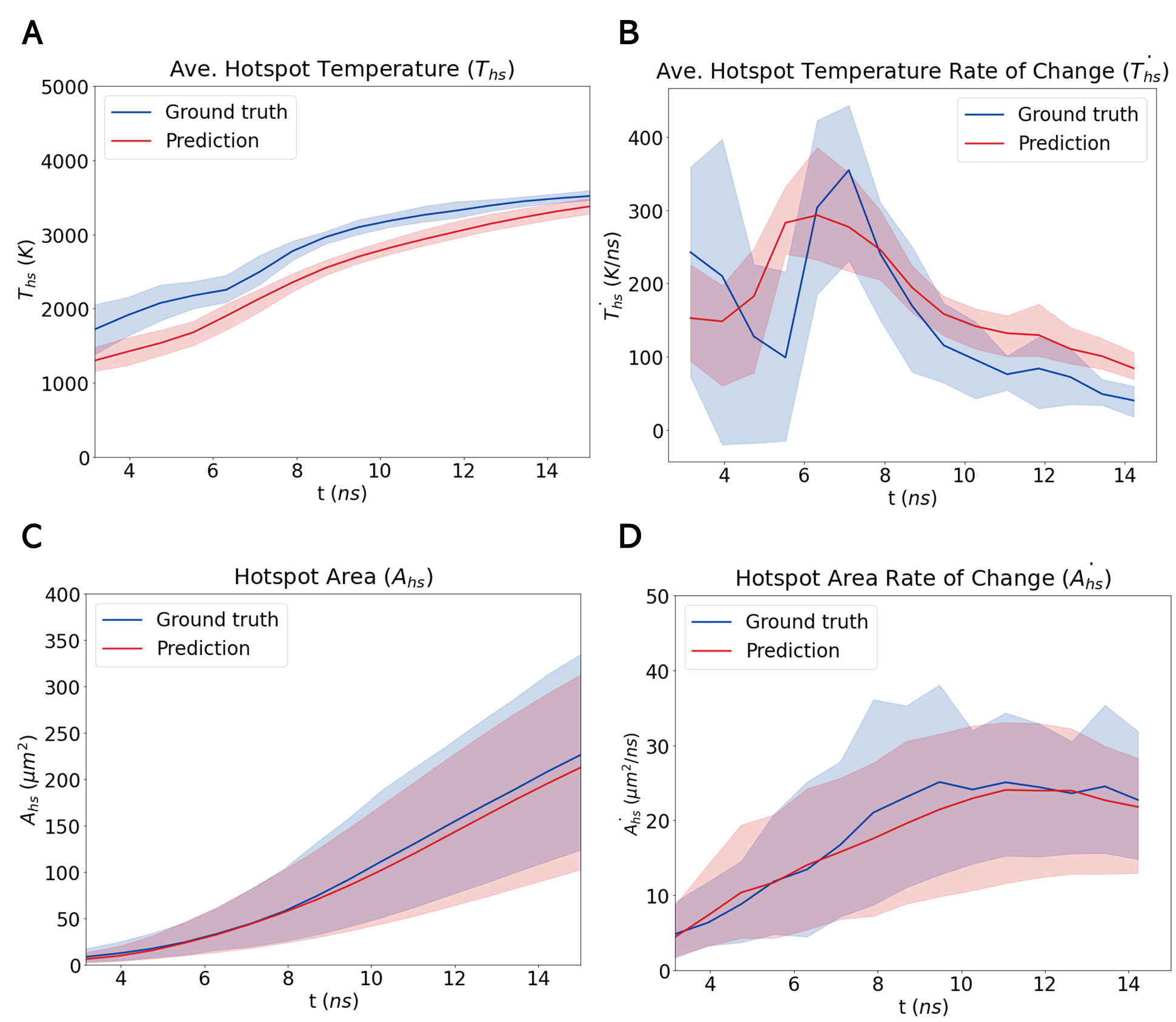}
    \caption{\textbf{EM sensitivity QoIs calculated from PARC predictions.} \textbf{(A)} Average hotspot temperature. \textbf{(B)} average hotspot temperature rate of change, \textbf{(C)} total hotspot area. \textbf{(D)} total hotspot area rate of change. Overall, these metrics are in good agreement with the DNS results, with the exception that the average hotspot temperature values are on average 352(K) lower in PARC predictions.}
        \label{fig:sensitivity}
\end{figure}

Figure~\ref{fig:sensitivity} illustrates the time evolution of the average hotspot temperature $\overline{T}^{hs}$ and total hotspot area ${A^{hs}}$ as well as their rate of change over time, $\dot{\overline{T}}^{hs}$ and $\dot{A}^{hs}$, derived from the PARC prediction illustrated in Figure~\ref{fig:PARC-result}. As depicted in Fig.~\ref{fig:sensitivity}, all four sensitivity QoIs derived from PARC agreed with those obtained from DNS at all observed time steps, with the exception of the average hotspot temperature, which was generally lower than values derived from DNS simulations.

We tested three statistical metrics, namely root mean squared error (RMSE), Pearson’s correlation coefficient (PCC) \cite{Mukhopadhyay2011}, and Kullback–Leibler divergence (KLD) \cite{Joyce2011}, between the PARC predicted sensitivity QoIs and the DNS generated values. The above metrics were derived from the sensitivity prediction results of all nine test samples. Details on the derivation of such metrics are given in the Supplementary Materials. As reported in Tab.~\ref{tab:sensitivity}, across the nine test samples, the RMSE of PARC prediction was low for all sensitivity QoIs, having the value of $374.09\ K$, $91.48\ K/ns$, $18.03\ \mu m^2$, and $3.92\ \mu m^2/ns$ for $\overline{T}^{hs}$,  $\dot{\overline{T}}^{hs}$, ${A^{hs}}$, and $\dot{A}^{hs}$, respectively. Especially, the PARC-predicted results for ${A^{hs}}$ and $\dot{A}^{hs}$ displayed low KLD (0.039 and 0.051, respectively), as well as high PCC (0.884 and 0.849, respectively), indicating strong statistical agreement. This, however, was not the case for the temperature QoIs ($\overline{T}^{hs}$,  $\dot{\overline{T}}^{hs}$). This indicates that PARC can successfully predict the evolution of hotspots that is consistent with DNS, with a caveat that hotspot temperatures are slightly underpredicted.

\begin{table}[bh!]
\centering
\begin{tabular}{|l|l|l|l|l|}
\hline
 & $\overline{T}^{hs}$ & $\dot{\overline{T}}^{hs}$ & $ {A^{hs}}$ & $\dot{A}^{hs}$ \\
\hline
RMSE & $374.09\ K$ & $91.48\ K/ns$ & $18.03\ \mu m^2$& $3.92\ \mu m^2/ns$\\
\hline
KLD & $8.377$ & $3.654$  & $0.039$& $0.051$ \\
\hline
PCC & $0.399$ & $0.131$ & $0.884$& $0.849$\\
\hline
\end{tabular}
\caption{\label{tab:sensitivity}\textbf{Evaluation of sensitivity QoI predicted by PARC}}
\end{table}

\subsection*{Computational Time}
Table~\ref{tab:computation_times} reports the average computational times for making a prediction using PARC, on different hardware configurations.  As reported, PARC was able to make predictions with only small computational times, approximately 0.5 seconds on a GPU-enabled system and 9.2 seconds on a CPU, which contrasts with hours of computational time for the DNS solver on multiprocessor HPC systems. As such, the small computational effort involved in PARC predictions makes it suitable for being used in microstructural design optimization in a materials-by-design framework.

\begin{table}[bh!]
\centering
\begin{tabular}{| p{2cm} | p{4cm} | p{4cm} | p{4cm} |}
\hline
& PARC (on CPU) & PARC (on GPU) & DNS \\
\hline
 Computation time & $\sim$ 9.2 seconds & $\sim$ 0.5 seconds & $\sim$ 24 hours \\
 (Average) & & & \\
\hline
 Hardware capacity & Intel® Core™ i9-11900 CPU @ 2.50GHz (16 cores; 1 was utilized for the experiment) & Intel® Core™ i9-11900 CPU @ 2.50GHz (16 cores; 1 was utilized for the experiment) &	Intel® Xeon E5-2699v4 CPU @ 2.80 GHz (1320 processors) \\
                    &       & & \\
                    & RAM: 64GB DDR4 & RAM: 64GB DDR4 & RAM: 128GB DDR4\\
                    &       & &\\
                    & GPU: N/A & GPU: Nvidia RTX A5000 & GPU: Nvidia P100 \\ 
\hline 

\end{tabular}
\caption{\label{tab:computation_times} \textbf{Computational efficiency comparison between PARC and DNS.} PARC drastically reduces the computation time (more than 9000 times) while requiring fewer computational resources.}
\end{table}

\subsection*{Physics-Awareness of PARC}
The physics-awareness is an essential trait of PARC that separates the model from conventional ``black-box'' ML models used in materials science. Currently, there are three common ways to achieve physics-awareness of an ML model, \textit{viz}., \textit{observational bias}, \textit{learning bias}, and \textit{inductive bias} approaches \cite{Karniadakis2021}. Observational bias approaches aim to embed physics into an ML model by training it with observational data reflecting underlying physical principles \cite{Yang2019b}. While straightforward to implement, observational bias approaches require a large amount of data to reinforce the physics bias and to produce predictions that follow the required physical laws \cite{Karniadakis2021}. Such a large data dependency makes them prohibitive to apply to many materials science applications, given that generating data via either numerical or physical experiments is generally expensive. On the other hand, learning bias approaches inject physics knowledge into an ML model via custom physics-based loss functions (training objectives) \cite{Raissi2019}. These approaches tend to be model-agnostic, as they rely only on the definition of loss functions, and therefore are more scalable and generalizable. However, as physics is embedded via soft constraints or penalties, these approaches can only produce a rough approximation of physical laws and, therefore, are not capable of producing accurate and high-fidelity results \cite{Karniadakis2021}. Finally, inductive bias approaches embed physics into an ML model by tailoring the mathematical formulation and the architecture of the ML model to reflect the underlying physics \cite{Long2019,Qin2019}. Similar to the way in which PARC models the governing differential equations of EM thermo-mechanics, inductive bias approaches attempt to express the underlying physical laws in the form of ML model parameters or architectural design \cite{Karniadakis2021}. Compared to the other two approaches, ML models designed in such a manner do not generally require a large quantity of data and can be trained with a small number of training samples (\eg only 30 simulation instances in the case of PARC; or 28 simulation instances in the case of Long \etal \cite{Long2019}), as the physics embedded in the model architecture and formulation serves as a strong prior for the learning process. Moreover, this method of embedding physics into ML models enables enhanced modeling capabilities that cannot be attained by physics-na\"{i}ve models. These modeling capabilities enabled by the physics-aware architecture will be further demonstrated in the following subsections.

\subsubsection*{Comparison with other physics-na\"{i}ve benchmarks}
To provide more quantitative evidence for the physics-awareness of PARC, we first investigated the roles of its differentiator and integrator networks by comparing PARC predictions to the results of other physics-na\"{i}ve models. In this experiment, two different baseline models were selected as benchmarks, namely the U-Net~\cite{Ronneberger2015} and the ImaGINator~\cite{Wang}. U-Net is a physics-na\"{i}ve version of PARC in which there are no recurrent differentiator-integrator layers. Note that the U-Net baseline is identical to the shape descriptor sub-network in PARC, which consists of 15,384,870 parameters, accounting for 86.18\% of the total number of parameters in PARC. Therefore, if PARC was not physics-aware and just simply \textit{memorized} how hotspots evolve based on different void shapes, a direct regression of the temperature and pressure fields from the U-Net outputs should have roughly the same accuracy as the PARC predictions. The second benchmark, namely the ImaGINator, is an archetype of a video generation algorithm in the computer vision community, which is capable of generating realistic videos of human motions. In brief, ImaGINator is a generative adversarial network (GAN) that is equipped with a spatiotemporal fusion mechanism that enables the generation of a realistic and smooth video sequence from a given image input. ImaGINator was originally designed to solve more advanced video generation problems, such as the generation of videos describing human facial expressions or human activities. These are problems that are orders of magnitude more difficult than the problem of generating the temperature and pressure fields. If the physics-awareness architecture of PARC has no influence on its prediction performance, then these more sophisticated algorithms should perform better, or at least on par, with PARC in predicting the hotspot evolution of shocked EM.

\begin{figure}[tb!]
    \centering
    \includegraphics[width=0.85\textwidth]{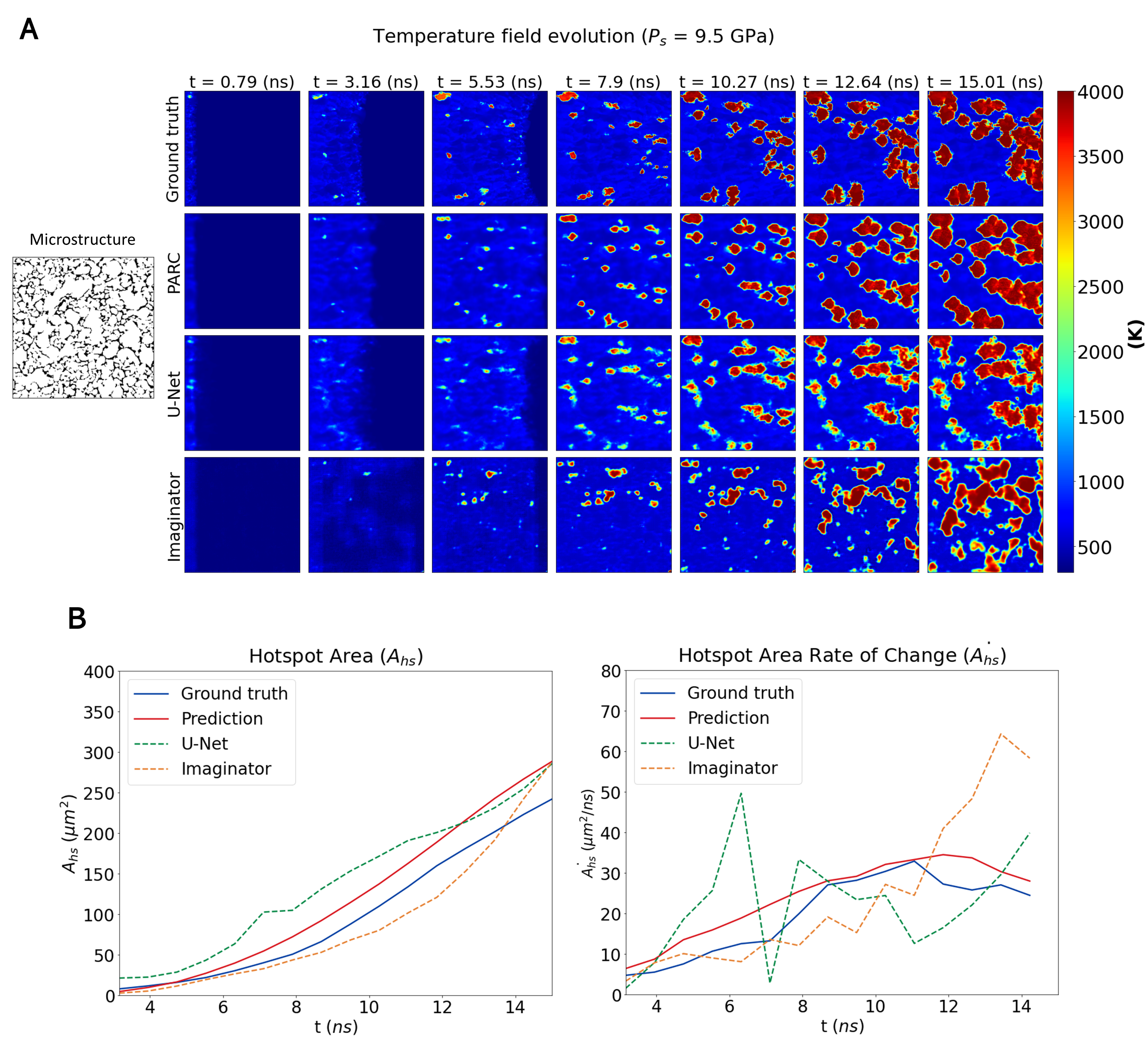}
    \caption{\textbf{Comparison between PARC and other physics-na\"{i}ve ML benchmarks.} \textbf{(A)} As illustrated in the temperature field predictions from PARC (\textit{second row}), U-Net \cite{Ronneberger2015} (\textit{third row}), and ImaGINator \cite{Wang} (\textit{last row}), in comparison with the DNS results (\textit{first row}), with its physics-aware architecture, PARC could predict hotspot formations and growths accurately, in contrast to the other two baselines. \textbf{(B)} The hotspot area (\textit{left}) and hotspot area growth rate (\textit{right}) predicted by PARC are also in better agreement with the ground truth DNS compared to other baselines. There is a discontinuous and fluctuated growth of hotspots area in the case of the U-Net baseline. Meanwhile the ImaGINator prediction is noisy and delayed compared to the ground truth DNS.}
    \label{fig:four_model_compare}
\end{figure}

Figure~\ref{fig:four_model_compare} (and Supplementary Figs.~\cref{fig:four_model_compare_1,fig:four_model_compare_2} for more samples) illustrates hotspot evolution prediction made by PARC in comparison with ones by the U-Net and Imaginator. Figure~\ref{fig:four_model_compare}A shows the temperature and pressure field predictions made by PARC and the two baseline models. Although the U-Net baseline was able to fairly accurately identify hotspot locations, the temperature values were considerably underestimated and the hotspot boundaries were blurry and fuzzy. More interestingly, the predicted hotspot growth was not smooth in the case of the U-Net baseline, rendering a noisy and discontinuous growth of the hotspot total area. Considering the thermo-mechanics, the growth rate of hotspots should be smooth and monotonic, as shown in the ground truth DNS and PARC predictions (Fig.~\ref{fig:four_model_compare}B). 

However, the U-Net baseline was better than PARC in the prediction of the shock profiles. As can be observed from Fig.~\ref{fig:four_model_compare}A, the U-Net baseline was able to predict the curved shape of the shock fronts, while PARC predictions were mostly flat and slightly shifted to the right. This, however, counterintuitively indicates that the U-Net baseline simply memorized the visual appearance of temperature fields at a given time step and was thus more effective in predicting shock profiles despite missing hotspot formation and growth. More closely, the shock loading condition was identical across all data samples, and thus the speed of the shock traveling through the microstructure and the resultant shock front shape were roughly consistent across all samples. On the contrary, hotspot formation and growth are complex functions of microstructures and their interactions. As a result, `one-to-one correspondence` memorizing approaches between a given microstructure and its corresponding temperature field evolution would be easily biased toward features that are common across training data samples (such as the shock profile) and would be incapable of producing good predictions of physically more critical and complex features (such as hotspot formation).

\begin{figure}[tb!]
\centering
\includegraphics[width=0.95\textwidth]{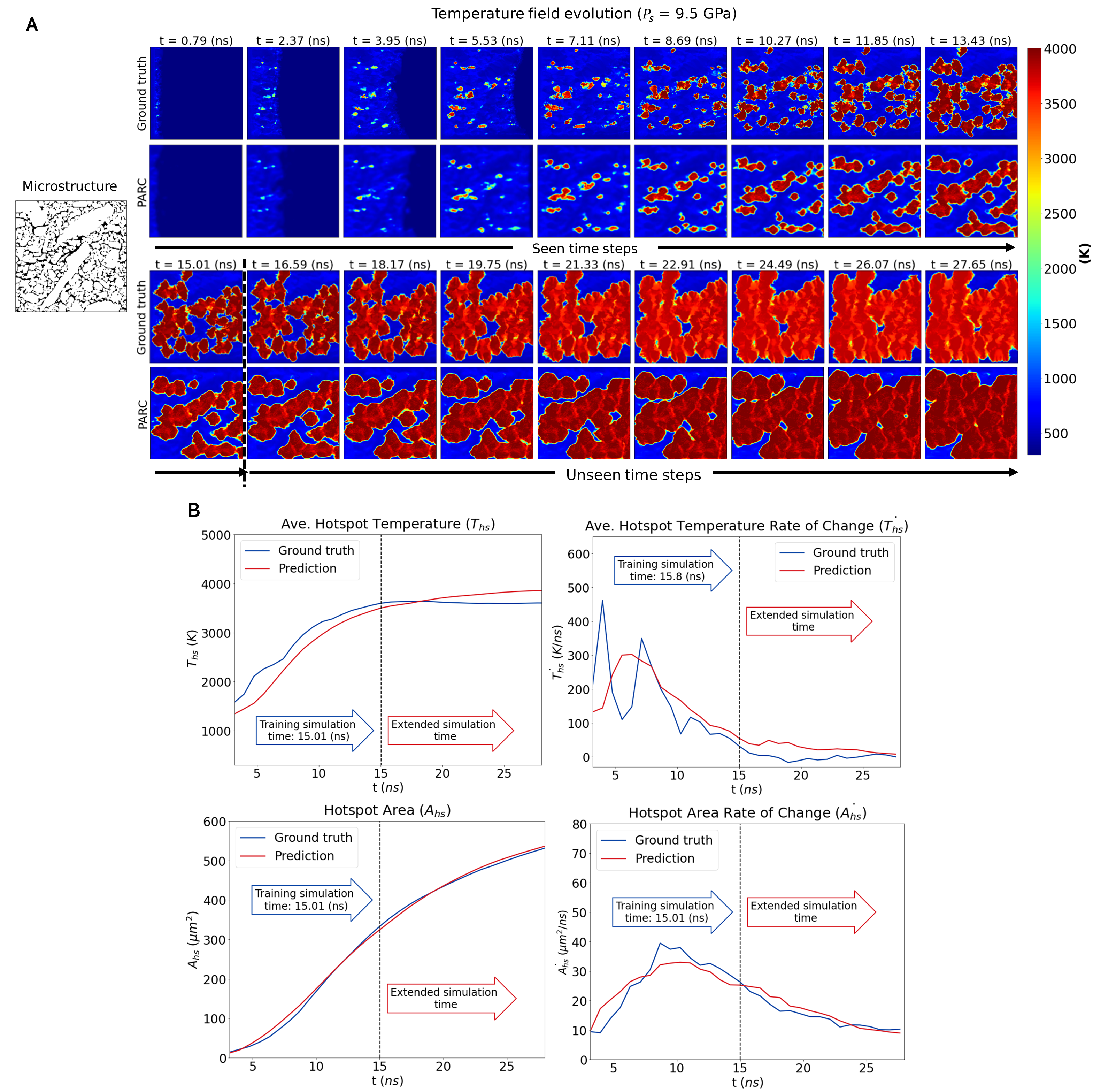}

\caption{\textbf{PARC predictions on unseen time steps.} \textbf{(A)} Temperature field evolutions in unseen time steps predicted by PARC. PARC had originally been trained for a total of 15.01 ns of simulation time. However, when we let it continue to predict time steps for a longer time frame, it was still able to predict the growth patterns well, although it reached the equilibrium state a little late and at a higher temperature. \textbf{(B)} Predicted sensitivity QoI in unseen time steps. PARC-predicted sensitivity QoI in unseen time steps also well agreed with DNS ground truth, despite the delayed and higher saturated temperature}
\label{fig:PARC-result-extended}
\end{figure}

The comparison between PARC and ImaGINator also supported the above observations. In theory, ImaGINator is a more advanced, state-of-the-art video prediction algorithm that can generate more sophisticated videos than what PARC can generate. However, ImaGINator was not able to predict the hotspot formations and growths anywhere near the accuracy that PARC could predict. The discontinuous shape of the shock fronts and the delayed and discontinuous hotspot area growth (Fig.~\ref{fig:four_model_compare}) clearly indicated that the physics-na\"{i}ve ImaGINator did not assimilate the thermo-mechanical principles of hotspot formation and growth.

In summary, the above benchmarking results demonstrate that PARC, with its physics-aware architecture, can achieve the enhanced modeling capabilities which are not feasible with generic, physics-na\"{i}ve ML models. This physics-aware architecture also enables PARC to generalize its knowledge to make predictions in previously unseen scenarios, as will be discussed in the following subsection.

\subsubsection*{Predictions extended beyond the time period of the training datasets}
To further attest the physics-awareness of PARC, we examined the model's ability to predict hotspot evolution beyond the time steps for which it was originally trained. Initially, PARC was trained for 19 time steps (which is equivalent to 15.01 ns). In this experiment, we allowed PARC to continue predicting the temperature and pressure fields for an additional 17 discrete time steps. Therefore, the extended simulation contained a total of 36 discrete time steps of shocked EM hotspot thermo-mechanics prediction (or 28.44 ns). Since PARC had not seen training examples beyond 19 time steps, we hypothesize that if PARC lacked physics-awareness and simply memorized the hotspot patterns at each time step, it would not be able to make physically meaningful predictions for future time steps.

Figure~\ref{fig:PARC-result-extended} (Supplementary Figs.~\cref{fig:PARC-result-extended-1,fig:PARC-result-extended-2} for more examples), shows that, in fact, PARC was able to maintain accurate predictions over these previously unseen time steps. Visually, the hotspot growth rate and pattern were in good agreement with the DNS results, although there was a delay in reaching the equilibrium state (\ie the plateauing of the hotspot temperature) and the equilibrium temperature was slightly higher than the DNS predicted values. Figure~\ref{fig:PARC-result-extended}B also confirms this inspection, showing good agreement between PARC predictions and ground truth DNS for all sensitivity metrics, even though the PARC-predicted average hotspot temperature is saturated at a later time and at a higher temperature. Thus, this experiment indicates that PARC has assimilated the physics of growth and coalescence of hotspots in the limited time period covered by the training dataset and is able to carry forward physically meaningful forward in time.

\begin{figure}[tb!]
\centering
\includegraphics[width=\linewidth]{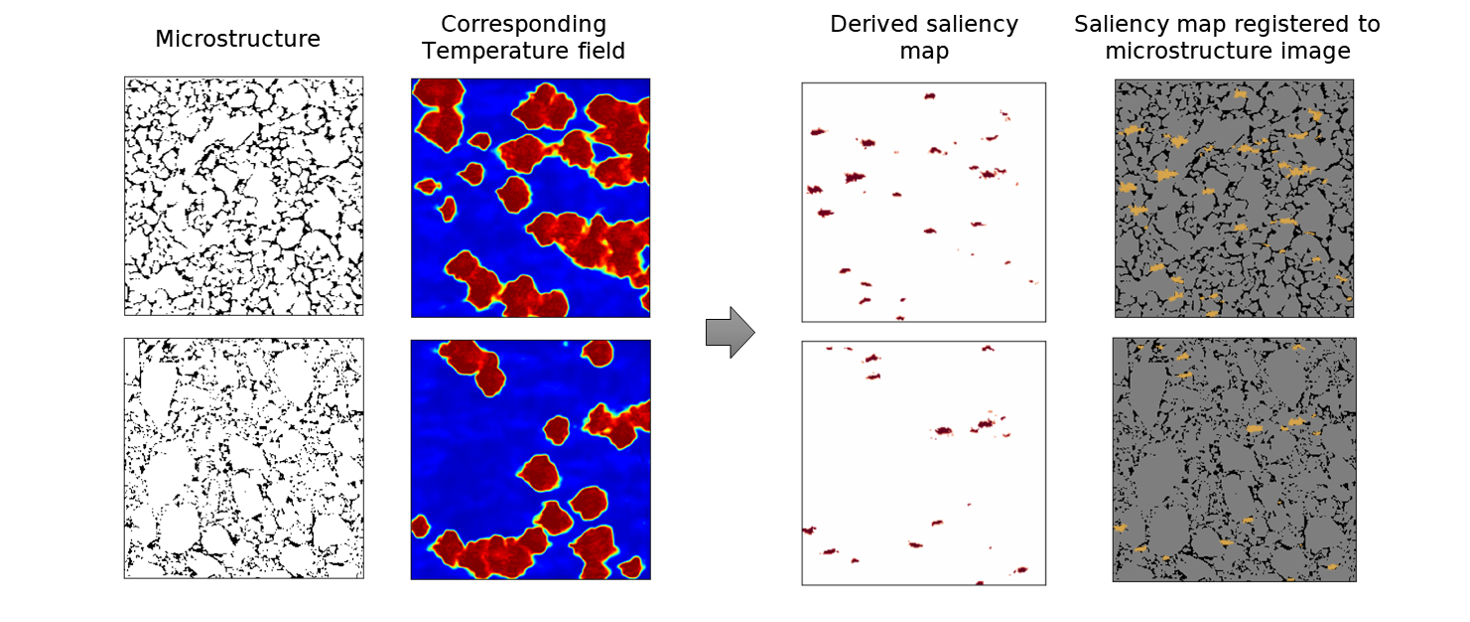}
\caption{\textbf{Saliency maps to visualize critical voids.} The saliency maps are derived by computing the derivative of corresponding temperature field \wrt input microstructure image. Consequently, the saliency map is registered to the corresponding microstructure image. The voids that are highlighted by the saliency map are classified as `critical.'}
\label{fig:saliency_map}
\end{figure}

\begin{figure}[tb!]
\centering
\includegraphics[width=0.9\textwidth]{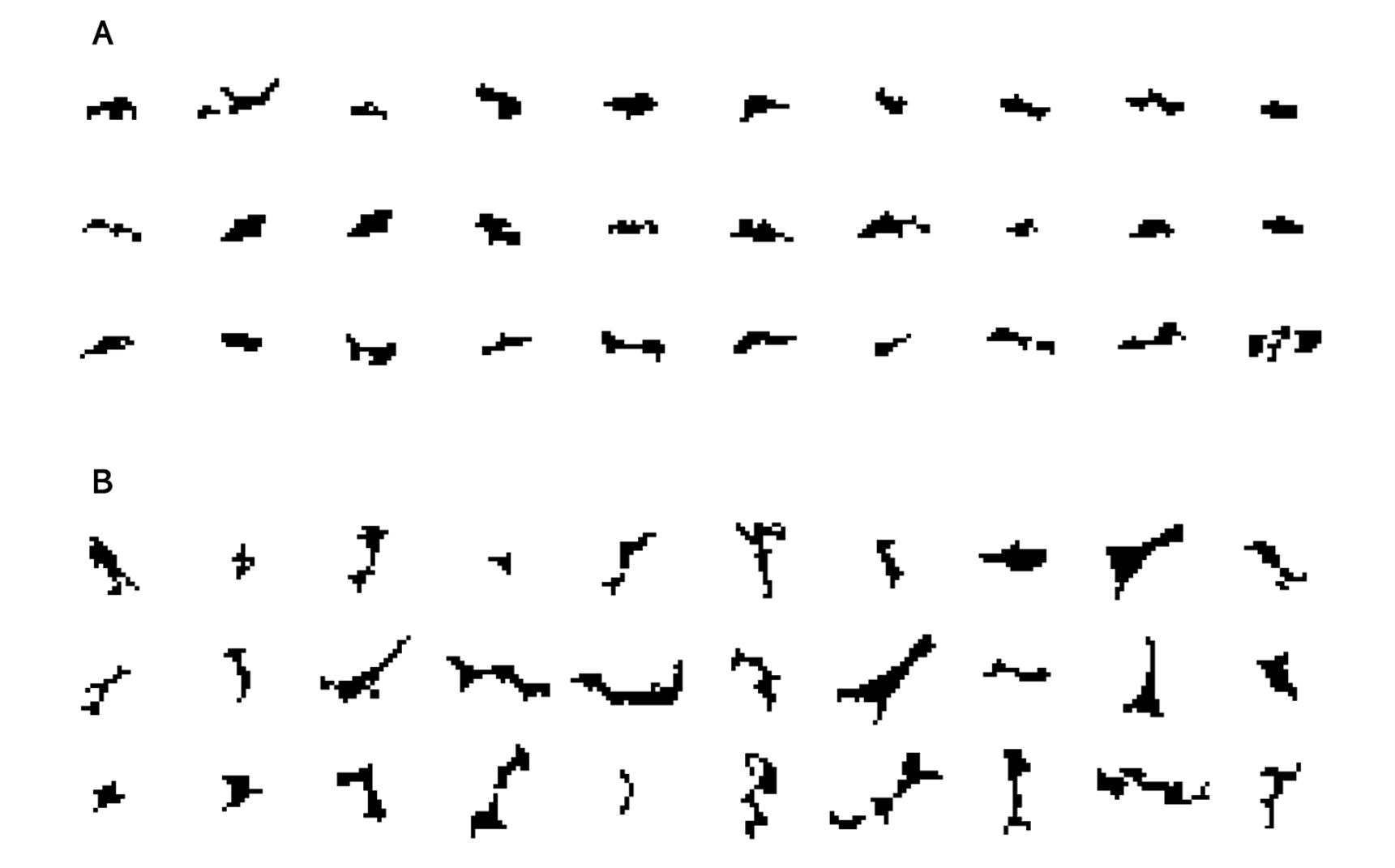}
\caption{\textbf{Examples of ``critical'' and ``non-critical'' voids identified from the saliency map visualization. (A)} ``Critical'' voids. \textbf{(B)} ``Non-critical'' voids.}
\label{fig:void_crop}
\end{figure}

\subsection*{Interpretability of PARC}
The physics-aware architecture of PARC also brings another advantage of being more interpretable compared to ``black-box'' ML approaches. As shown in Fig.~\ref{fig:PARC-general}, PARC incorporates a U-net-encoded microstructure shape descriptor, $\mu$, for the prediction of the time derivatives of temperature and pressure fields. This physics-aware architecture of PARC allows us to understand the effect of microstructure morphology on the formation of hot spots by employing the \textit{saliency map} technique \cite{Simonyan14a}. As explained in the Supplementary Materials, the saliency map visualizes regions in the microstructure that make a predominant contribution to the prediction of high-temperature regions (hotspots). PARC can identify microstructural features within such regions as `critical' and quantify the spatial features in the microstructure that lead to energy localization. Figure~\ref{fig:saliency_map} highlights microstructural features with high influence on the hot spot formation as visualized by employing the saliency map. The saliency map shows that the regions highlighted by high saliency values are colocated with a subset of the microstructure's voids and cracks. This finding is in line with prior knowledge in the EM research community \cite{Johnson1985,Massoni1999,Frey1982,Menikoff2004} that hot spots are formed due to the energy localization at sites of cracks and voids in EM microstructures.

\begin{figure}[tb!]
\centering
\includegraphics[width=\linewidth]{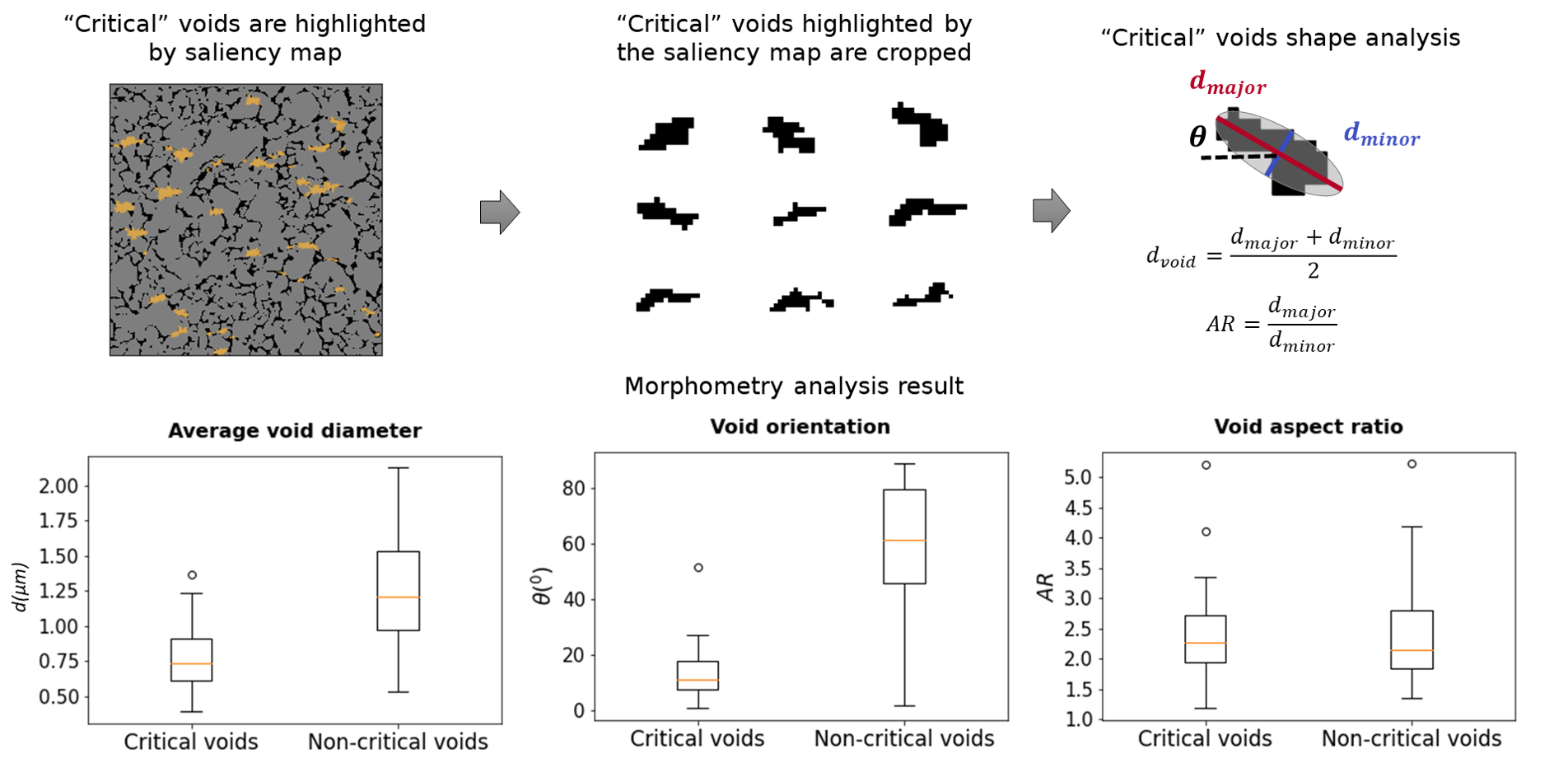}
\caption{\textbf{Morphometry analysis of ``critical'' voids identified by PARC.} The result shows a clear distinction in the average void diameter and the void orientation between ``critical'' and ``non-critical'' voids. Their \textit{p}-values were $8.6\times10^{-12}$ and $2.7\times10^{-22}$, respectively. Meanwhile, the distinction in the void aspect ratio was statistically insignificant ($p=0.9989$).}
\label{fig:morphometry_analysis}
\end{figure}

To further highlight the physical implications of the saliency map visualization, we analyzed the microstructural features that lead to critical hotspot formation. We first classified the voids and cracks in the microstructure into `critical' and `non-critical' ones and analyzed the shape of both groups. Figure~\ref{fig:void_crop} shows some samples of critical and non-critical voids, randomly selected from the saliency map visualization results. We found that most of the critical voids were elongated in shape, and the predominant orientation of these voids was parallel to the direction of shock propagation (from left to right in the domain), as can be surmised by visual inspection of the samples in Fig.~\ref{fig:void_crop}A. These findings align with the observations in the prior works of Roy \etal \cite{Roy2020a}, Rai and Udaykumar \cite{Rai2015}, Nguyen \etal \cite{Nguyen2022}, and others
, which demonstrated the strong influence of void aspect ratios and void orientations on the formation of hotspots. 

To further elucidate this insight gained from the saliency map, we fit an effective ellipsoid to these voids and cracks to compute the void size, aspect ratio, and orientation (Fig.~\ref{fig:morphometry_analysis}). This morphometry analysis showed that voids with an average diameter of $0.761\ \mu m\ (\pm0.224)$, aspect ratio of $2.37\ (\pm0.69)$ and angle of inclination of $12.752^{\circ}\ (\pm8.699)$ \wrt the shock direction were most likely to form critical hotspots, whereas, non-critical regions of the microstructures had the average void diameters of $1.247\ \mu m\ (\pm0.383)$, aspect ratio of $2.373\ (\pm0.826)$, and orientation of $59.192^{\circ}\ (\pm24.49)$ \wrt to the shock direction. Two-sample t-test\cite{Seltman2012} also confirmed the clear differences in the morphometry distributions of critical and non-critical regions with small \textit{p}-values of $8.6\times{10}^{-12}$ and $2.7\times{10}^{-22}$ for the average void diameter and orientation, respectively, indicating their statistically meaningful tendencies. Meanwhile, there was no clear distinction in the void aspect ratio between critical and non-critical voids, as the two-sample t-test resulted in high \textit{p}-value of $0.9959$. 

The above morphometry analysis again agrees with the previous findings on the effect of elongated void orientation on the hot spot formation using DNS, \ie that voids with their major axis aligned parallel to the direction of shock propagation are the potential sites for energy localization \cite{Rai2015}; on the other hand, voids oriented nearly perpendicular to the shock direction fail to produce sustained hotspots \cite{Nguyen2022}. Also, the analysis result also agrees with the conclusion of Nguyen \etal \cite{Nguyen2022} that the orientation of voids has a stronger impact on the energy localization that causes hotspot ignition compared to their aspect ratio. This agreement between the analysis provided by the saliency map of PARC and previous DNS-based physical insights demonstrates the possibility of examining the PARC neural network to inform EM designers regarding the structure-property-performance linkages for EMs.

\section*{Discussion}
This work presented a deep-learning approach to assimilate the thermo-mechanics of shock-initiated heterogeneous EMs with complex microstructures. The PARC neural network model was developed based on the recurrent differentiator-integrator architecture, which resembles the way in which physics equations are modeled and solved in numerical simulations. The PARC model was then trained using grid-collocated field data in the form of discrete video frames produced by DNS of hotspot evolution in shocked microstructures. Following training, PARC developed the capability to predict the hotspot dynamics for an unseen microstructure that was not included in the training dataset.

The predictive capability of PARC was validated against DNS in two classes of pressed energetic material (HMX) microstructures, \textit{viz}., Class V and fluid-energy-milled (FEM) materials. The results showed that PARC could deliver high-fidelity and accurate predictions of temperature and pressure field evolution in shocked heterogenous EM microstructures. In addition, the PARC-predicted EM sensitivity QoIs, including hotspot temperature, hotspot area, and their rate of change over time, showed good agreement with those derived from the ground truth DNS. The computational cost of employing PARC was multiple orders of magnitude lower than the case of DNS. PARC could produce prediction results in compute times in the order of a few seconds on a standard laptop, while DNS occupied hours to days on multiprocessor machines. Therefore, while it is infeasible to employ DNS as part of a concurrent multiscale modeling framework to connect sub-grid (mesoscale) dynamics with macroscale hydrocodes, PARC opens the possibility of intercalating macro-calculations with on-the-fly training and interrogation of the mesoscale dynamics.

In addition, the physics-awareness of PARC was demonstrated through multiple experiments. Compared with other physics-na\"{i}ve baselines, PARC showed a better and more reliable performance. Especially, the comparison against the U-Net baseline implied that PARC did not just memorize the correspondences between microstructural elements (\ie voids) and hotspot formation but assimilated the underlying thermo-mechanics of hotspot formation in shocked EM. Similarly, the benchmarking against a generic, state-of-the-art video generation network called ImaGINator revealed that the physics-aware architecture of PARC enabled the prediction results to be physically plausible. Furthermore, when challenged to predict scenarios beyond what it had originally been trained for, PARC was able to make physically meaningful and reliable predictions. All of these results are indicative of the physics-awareness of the differentiator-integrator architecture of PARC.

Furthermore, we showed that the physics-aware architecture of PARC can circumvent the pitfalls of ``black-box'' machine learning approaches and enable better modeling capabilities and generalization. Specifically, through the quantification of the saliency map, PARC enabled the establishment of correlations between morphological features and reaction sensitivity. For instance, the indication from PARC that elongated voids with their major axis aligned parallel to the direction of shock propagation present the most favorable sites for critical hotspots is in good agreement with prior expert knowledge derived from direct numerical simulations. Furthermore, the analysis of the saliency map also indicated that the orientation of elongated voids with respect to the incident shock has the strongest influence on the criticality of voids compared to other morphometry parameters such as void size and void aspect ratio.

We are currently undertaking to extend the capability of PARC in several directions. One of the immediate challenges is to increase PARC prediction accuracy to capture details such as the shape of the shock structure passing through the EM microstructure and the wave propagation patterns in the pressure field. Additionally, the present work is also limited to a single operating condition (shock strength of 9.5 GPa), whereas there is a clear practical need for generalizing the results to other operating conditions.

In the long term, PARC can facilitate the materials-by-design of EMs by accelerating the determination of structure-property-performance (SPP) linkages. With the ability to incorporate non-idealized microstructures and provide rapid predictions of shock-initiated EM thermo-mechanics, PARC can accelerate the design optimization process and shorten the time to discover EM microstructures that provide tailored performance characteristics. This warrants further investigation into how PARC may be integrated into the design process of EMs.

Furthermore, the interpretability of PARC and the corresponding visualization techniques may provide some additional lenses that can be used to expand our physical knowledge connecting morphology to performance. While the present study is limited to saliency visualization, different neural network visualization methods for explainable machine learning may reveal interesting, previously uncovered correlations between structure, property, and performance. To this end, exploring more evocative neural network visualization techniques is another avenue for further investigation.

Finally, although the scope of the present work was limited to EMs, the methods developed can be applied to other complex materials with a strong microstructural influence on properties and performance. Hence, the generalization of PARC to a wider range of materials may yield similar benefits, such as accelerating the discovery of materials with enhanced properties via microstructure design. Conversely, different material problems may pose different physical constraints (\eg rotational symmetry, conservation of energy), which may lead to the development of new mathematical foundations and algorithms for deep learning, rendering an opportunity for multidisciplinary research.

\section*{Materials and Methods}

PARC was trained on a data set containing 42 instances of shock-initiated reaction simulations on two different classes of pressed EMs: Class V (Supplemental Fig.~\ref{fig:prob}A) and fluid-energy-milled (FEM; Supplemental Fig.~\ref{fig:prob}B). The microstructures of these two materials were obtained from scanning electron microscope (SEM) images with spatial dimensions of $25\ \mu m\times25\ \mu m$ resolved with $240\times240$ pixels ($104$ nanometers per pixel). 

Sequences of temperature and pressure fields evolving over time were computed from DNS using the multi-material reactive dynamics code, SCIMITAR3D \cite{Rai2015,Rai2017b,Rai2018}. The numerical calculations were performed using methods and models which are not detailed here in the interest of brevity but are described in several publications \cite{Rai2017a,Rai2020b,Das2020}. For each DNS, the microstructural sample was loaded with a shock with the pressure loading of 9.5 GPa, applied at the left boundary of the domain. The shock then traversed through the microstructure from the left to right (Supplemental Fig.~\ref{fig:prob}C). The total time duration of each simulation was $15.01$ nanoseconds $(ns)$, which was sufficient for the shock to traverse the entire sample and for the collapse of all voids in the material. The time evolution of hotspots in a typical DNS can be visualized in the video files in the Supplementary Materials. Hotspot ignition and growth ensued in the period of $15.01$ ns, creating a field of hotspots that evolve in the post-shock region. The calculated temperature and pressure field data were recorded at equal time intervals leading to 19 equally-spaced time instances at $\Delta t=0.79$ ns time interval. 

During the above numerical simulations, the initial microstructures were deformed and advected with the post-shock flow velocity. To relate the initial void field in the microstructure to the final hotspot field in the fixed grid framework, the DNS temperature and pressure fields were pre-processed so that the hotspot evolution was correlated to the initial voids in the microstructure. This was accomplished using a backtracking procedure using the algorithm described in the Supplementary Materials. Briefly, the backtracking procedure utilizes the known (DNS) velocity field at each snapshot of the temporal evolution and recursively maps the hotspot field back to the initial configuration of the microstructure. The Supplementary Materials show videos of the evolving microstructure and the resulting hotspot fields following the backtracking process for one example case. 

The time-evolving temperature and pressure fields were normalized to produce values at each pixel location ranging from -1 to 1. The original data set of 42 samples was split into training, validation, and testing sets, with 30, 3, and 9 samples, respectively. The neural network parameters were initialized using the normalized He initialization method \cite{He2015} and the model was trained using the ADAM optimizer \cite{Kingma2015} with the learning rate of ${10}^{-4}$. The codes for training PARC and predicting EM hotspot dynamics, as well as generating the figures presented in this paper, may be found online at \url{https://github.com/stephenbaek/parc}.

\section*{Acknowledgements}
We are grateful to the reviewers who inspired the experiments regarding the physics-awareness of PARC.

\paragraph{Funding}
This work was supported by the U.S. Air Force Office of Scientific Research (AFOSR) Multidisciplinary University Research Initiative (MURI) program (Grant No. FA9550-19-1-0318; PM: Dr. Martin Schmidt, Dynamic Materials program), and partially based upon work supported by the National Science Foundation under Grant No. 2203580.

\paragraph{Author contributions}
S.B. conceived the idea and developed it with H.U. S.B. implemented PARC originally and P.N. further developed the code. P.N. conducted the experiments with the assistance of J.C. Y.N. and P.S. produced DNS results. P.N. and Y.N. created visualizations. S.B. and H.U. supervised the project. All authors contributed to analyzing the results and writing the manuscript.

\paragraph{Competing Interests}
The authors declare that they have no competing interests. 

\paragraph{Data Availability Statement}
All data needed to evaluate the conclusions in the paper are present in the paper and/or the Supplementary Materials. All code and data can be downloaded from the permanent repository \url{https://dx.doi.org/10.5281/zenodo.7309101}, as well as a GitHub repository maintained by the authors \url{https://github.com/stephenbaek/parc}.

\bibliography{main}

\bibliographystyle{Science}

\newpage
\appendix
\renewcommand{\thefigure}{S.\arabic{figure}}    
\renewcommand{\theequation}{S.\arabic{equation}}

\setcounter{equation}{0}
\setcounter{figure}{0}

\section*{Supplementary Materials}
\subsection*{Additional Details on the PARC Architecture Design}
\begin{figure}[htb!]
\centering
\includegraphics[width=0.7\textwidth]{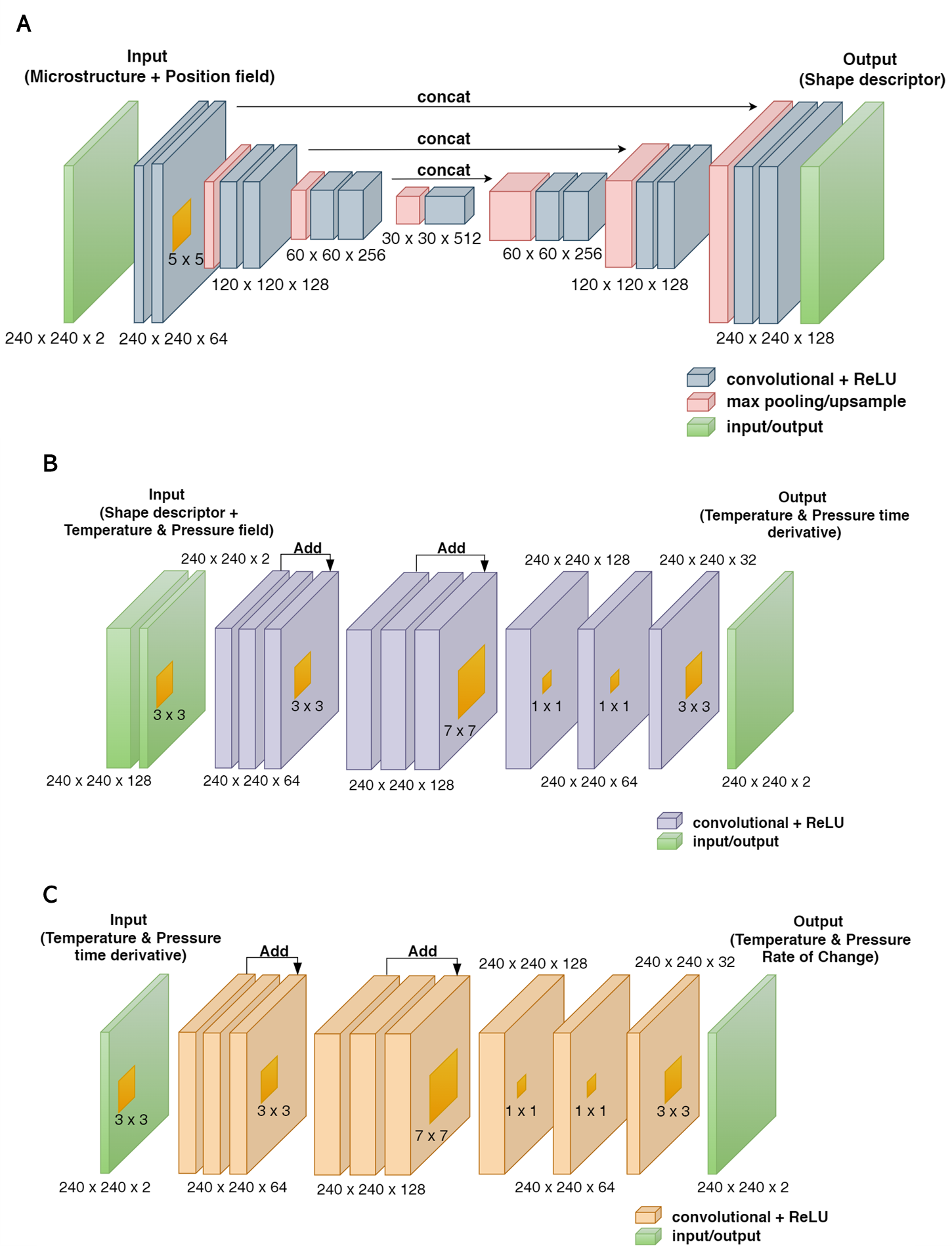}
\caption{\textbf{Architecture design of PARC. (A)} The U-Net shape descriptor extractor network. \textbf{(B)} Differentiator network. \textbf{(C)} Integrator network.}
\label{fig:PARC_arch}
\end{figure}

\paragraph{U-Net for Shape Descriptor Extraction}
The U-Net is an encoder-decoder-based neural network that is widely used for semantic segmentation tasks \cite{Ronneberger2015}. As illustrated in Fig.~\ref{fig:PARC_arch}A, the U-Net employed in this work is comprised of two network paths: encoder (compression) and decoder (expansion). The encoder takes a $240\times240\ (spatial)\times1$ grayscale microstructure image $I(x,y)$ and a $240\times240\times1$ position map $U(x,y)$ as inputs. First, two convolution layers with the rectified linear unit (ReLU) activation produce $240\times240\times64$ low-level feature maps. Consequently, the low-level feature maps are then down-sampled using the $2 \times 2$ max-pooling operation with the stride of two, in order to reduce the spatial dimensions of the feature maps by half and thereby double the size of the receptive field. The down-sampled feature maps are processed through two subsequent convolutional layers, resulting in $120\times120\times128$ feature maps. Kernel sizes are $5\times5$ for all convolutional layers. A similar process is applied two more times, such that at each time the spatial dimensions of feature maps reduce by half and the number of feature maps increases twice, resulting in throughputs of the size $60\times60\times256$ and $30\times30\times512$, respectively. The final $30\times30\times512$ feature map is a high-level abstraction representation of the original microstructure image. The high-level feature map is then reconstructed by the decoder whose architecture is symmetric to the encoder, to produce a map of shape descriptors that can be used in the derivative network. Unlike the conventional use of U-nets, the U-Net subnetwork in PARC outputs a 128-channel feature map, containing shape descriptors across different image locations. Enabled by the multi-level, multi-resolution architecture of the U-Net, these shape descriptors codify local morphological characteristics of the microstructure in a hierarchical manner.

\paragraph{Differentiator and Integrator CNNs}
As illustrated in Figs.~\ref{fig:PARC_arch}B and \ref{fig:PARC_arch}C, the architectures of the differentiator and the integrator are identical, except for the input layers. The differentiator network has $240\times240\times130$ input dimensions constituted by $\mathbf{X}\in\mathbb{R}^{240\times240\times2}$, the temperature and pressure fields collocated on the simulation grid, and $\mu\in\mathbb{R}^{240\times240\times128}$, the microstructure shape descriptor. The output of the differentiator is the time derivative $\partial\mathbf{X}/\partial t$ of the state $\mathbf{X}$, which then feeds the integrator. Hence the input to the integrator has the dimensions of $240\times240\times2$. Aside from the input layers, both of the networks are comprised of two ResNet blocks \cite{He2016} and a super-resolution block \cite{Dong2015} of the identical architecture. The first ResNet block contains three $3\times3$ convolution layers with zero padding to produce a $240\times240\times64$ feature map. The ResNet skip connection adds the output of the first convolution layer to that of the last convolution layer of the ResNet block. The second ResNet block is the same as the first ResNet block, except for having 128 channels in the output. The super-resolution block starts with a $7\times7$ convolution layer to produce a $240\times240\times128$ feature map, followed by two $1\times1$ convolution layers to produce $240\times240$ feature maps with the number of channels 64 and 32, respectively. The final output of the SR block is then processed through a $3\times3$ convolution layer, producing the final output of the dimension $240\times240\times2$.

\subsection*{Evaluation Metrics for the Sensitivity QoIs}

The accuracy of PARC-predicted the average hotspot temperature, total hotspot area, and their rates of change over time is evaluated via several metrics, including root mean squared error (RMSE), Pearson’s correlation coefficient (PCC) \cite{Mukhopadhyay2011}, and Kullback–Leibler divergence (KLD) \cite{Joyce2011}:

\begin{equation}
  \text{RMSE}(x, \hat{x}) := \sqrt{\frac{1}{N \times P}\sum_{l=0}^{N}\sum_{k=0}^{P}\left[x_l\left(t_k\right)-\hat{x_l}\left(t_k\right)\right]^2},
\end{equation}

\begin{equation}
  \text{PCC}(x, \hat{x}) := \frac{1}{P}\sum_{k=0}^{P}\frac{1}{\sigma(t_k)\hat{\sigma}(t_k)}\sum_{l=0}^{N}\left[x_l\left(t_k\right)-\mu\left(t_k\right)\right]\left[\hat{x_l}\left(t_k\right)-\hat{\mu}\left(t_k\right)\right],
\end{equation}

\begin{equation}
  \text{KLD}(x, \hat{x}) := \frac{1}{P}\sum_{k=0}^{P}\left[\log{\frac{\sigma\left(t_k\right)}{\hat{\sigma}\left(t_k\right)}}+\frac{\hat{\sigma}\left(t_k\right)-\left\{\hat{\mu}\left(t_k\right)-\mu\left(t_k\right)\right\}^2}{2\sigma^2\left(t_k\right)}-\frac{1}{2}\right].
\end{equation}

\noindent Here, $x$ and $\hat{x}$ denote a sensitivity QoI derived from PARC and DNS, respectively, and $\mu$ and $\sigma$ represent the mean and the standard deviation of the sensitivity QoI, with the similar use of the hat to distinguish PARC and DNS derived values. $P$ is the number of total time steps and $N$ is the number of testing samples.

\subsection*{Saliency Map}
Saliency maps \cite{Simonyan14a}, or pixel attribution methods, are a way of visualizing CNNs by highlighting image areas that played important roles in the prediction of the outputs. In this paper, we first compute the derivative of the predicted temperature fields \wrt the individual pixel values of the input microstructure image, to measure the attributions of microstructural elements to the prediction of the temperature field:
\begin{equation}
\label{eqn:saliency_single}
  \mathbf{G}\left(t_k\right)=\ \frac{d\mathbf{T}\left(t_k\right)}{d\mathbf{I}}.
\end{equation}
We then compute the sum of such derivatives over all time steps, to account for the time-evolving temperature values:
\begin{equation}
  \mathbf{G}= \sum_{k=1}^{P}{\mathbf{G}(t_k)}.
\end{equation}
The raw saliency map produced in this manner can be fuzzy and difficult to interpret. Hence, we apply a threshold value $\varepsilon$ to filter out the areas with small or negative gradient values during the backpropagation. The threshold value is selected empirically.
\begin{equation}
  \left\{
    \begin{array}{ccccc}
        G_{ij} &= & G_{ij} &if &G_{ij}\geq \varepsilon ,   \\
        G_{ij} &= & 0 &otherwise .   
    \end{array}
  \right.
\end{equation}

\subsection*{Hotspot Ignition and Growth Simulations}
The shock-induced energy localization in EM occurs via several mechanisms, including void collapse, plastic dissipation, and intergranular or interface friction. However, under the high shock strength, void collapse is the major source of energy localization while plastic dissipation and inter-granular/interface friction play a minor role in the formation of hotspots. Therefore, the present work focuses on void-collapse-induced hotspot formations.

Mesoscale simulations are performed to simulate the collapse of voids in the microstructure due to the passage of a shock wave. SCIMITAR3D, an in-house multi-material flow solver is used for all simulations. It is a well-tested and validated solver for computing the reactive shock dynamics of energetic materials \cite{Rai2015,Rai2017b,Rai2018} and employs a Cartesian grid-based sharp-interface framework for compressible multi-material flows \cite{Kapahi2013}. The governing equations for the motion and deformation of the EM are cast in an Eulerian form \cite{Kulikovskii1999}. The numerical framework has been described in detail in previous works\cite{Kapahi2013,Rai2017a,Rai2020b,Das2020} and validated against experiments \cite{Rai2020b} and molecular dynamics simulations \cite{Das2021} for high-speed multi-material shock and impact problems. Details of the numerical treatment can be obtained from the above-cited previous works. Notably, in this work, we study a neat-pressed energetic material which only contains solid HMX phase and void phases in the materials.

To solve the reactive shock dynamics of the pressed HMX system in the present work, the CT images of pressed HMX microstructures are input to the simulation using image processing approaches described in previous work \cite{Dillard2014}. The interfaces between the crystal and void are embedded in the Cartesian grid and tracked using the level-set method 
and the ghost fluid method (GFM) 
is used to impose appropriate boundary conditions at the material-material or material-void interfaces. 
An adaptive local mesh refinement scheme is employed to adequately resolve shocks, reaction fronts and interfaces.
The energetic material HMX is modeled as a reactive material employing a 3-equation reduced-order reaction model \cite{Tarver1996}. Note that the material models for HMX are the most up-to-date ones currently available in the literature \cite{Das2021}. Among the available reaction chemistry models for HMX, the Tarver 3-equation model has been shown to most closely reproduce experimental data on criticality curves for neat pressed HMX microstructures in recent work \cite{Roy2020b}. 

Reactive calculations are performed using methods discussed extensively in previous works \cite{Rai2015}. The temperature, pressure, and species field data output from the direct numerical simulations are utilized to quantify the response of the pressed material to the imposed shock. In the present context, the quantities of interest (QoIs) for the effect of microstructure on the sensitivity of the material are the evolution of the temperature field and the reaction product mass fraction in the sample. The temperature field $T\left(\mathbf{x},t\right)$ in the domain measures the intensity of a hot spot resulting from the process of void collapse. Higher temperature hot spots formed due to the collapse of voids in the material lead to high localized energy release rates. The reaction product mass fraction of HMX, \ie the total mass of solid HMX material converted to gaseous species at any time t, is used to quantify the physio-chemical response of the material and is given by the following equation:

\begin{equation}
\label{eqn:react}
  F\left(t\right)=\frac{M_{reacted}\left(t\right)}{M_{HMX}}
\end{equation}

In Eq.~\ref{eqn:react}, $M_{HMX}$ is the mass of HMX in the total sample prior to the beginning of the chemical reaction process in the material, while $M_{reacted}$ is the mass of complete reaction products resulting from the burning of solid HMX to result in final gaseous reaction products. Complete conversion of the solid HMX to product gases is reached when $F=1$. The reaction zone defines the hotspot in the domain, which is defined as the region where the temperature of the material exceeds the value of the temperature ($T_{bulk}$) reached after the passage of a planar shock wave. The hot spot area $A_{hs}$ is an important QoI for determining sensitivity and is calculated as the area of the domain where the temperature $T\left(\mathbf{x},t\right)>T_{bulk}$. The hot spot area is recorded throughout the simulation to track the evolution of the hot spots.

\begin{figure}[tb!]
\centering
\includegraphics[width = 0.6\textwidth]{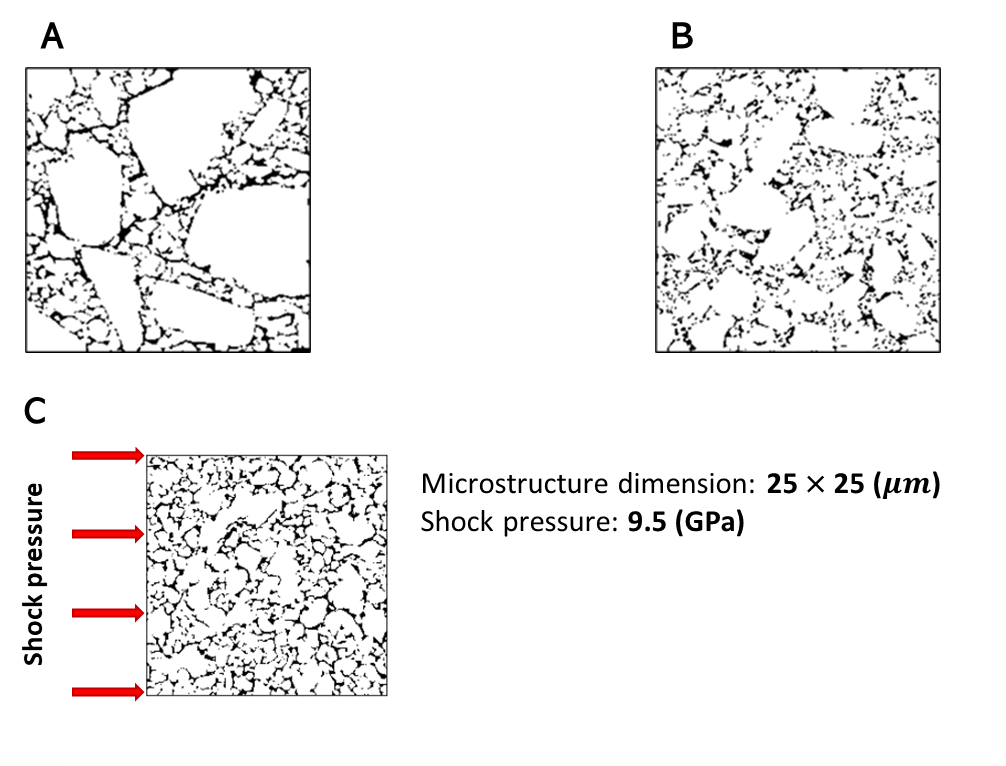}
\caption{\textbf{HMX microstructures and the loading condition of the shock-initiated simulation. (A)} Class V microstructure. \textbf{(B)} FEM microstructure. \textbf{(C)} The shock-initiated simulation is conducted with a shock pressure of 9.5 GPa entering the EM microstructure with the dimension of $25\times25\ \left(\mu m\right)$ from the left.}
\label{fig:prob}
\end{figure}

\begin{figure}[tb!]
    \centering
    \includegraphics[width=0.75\textwidth]{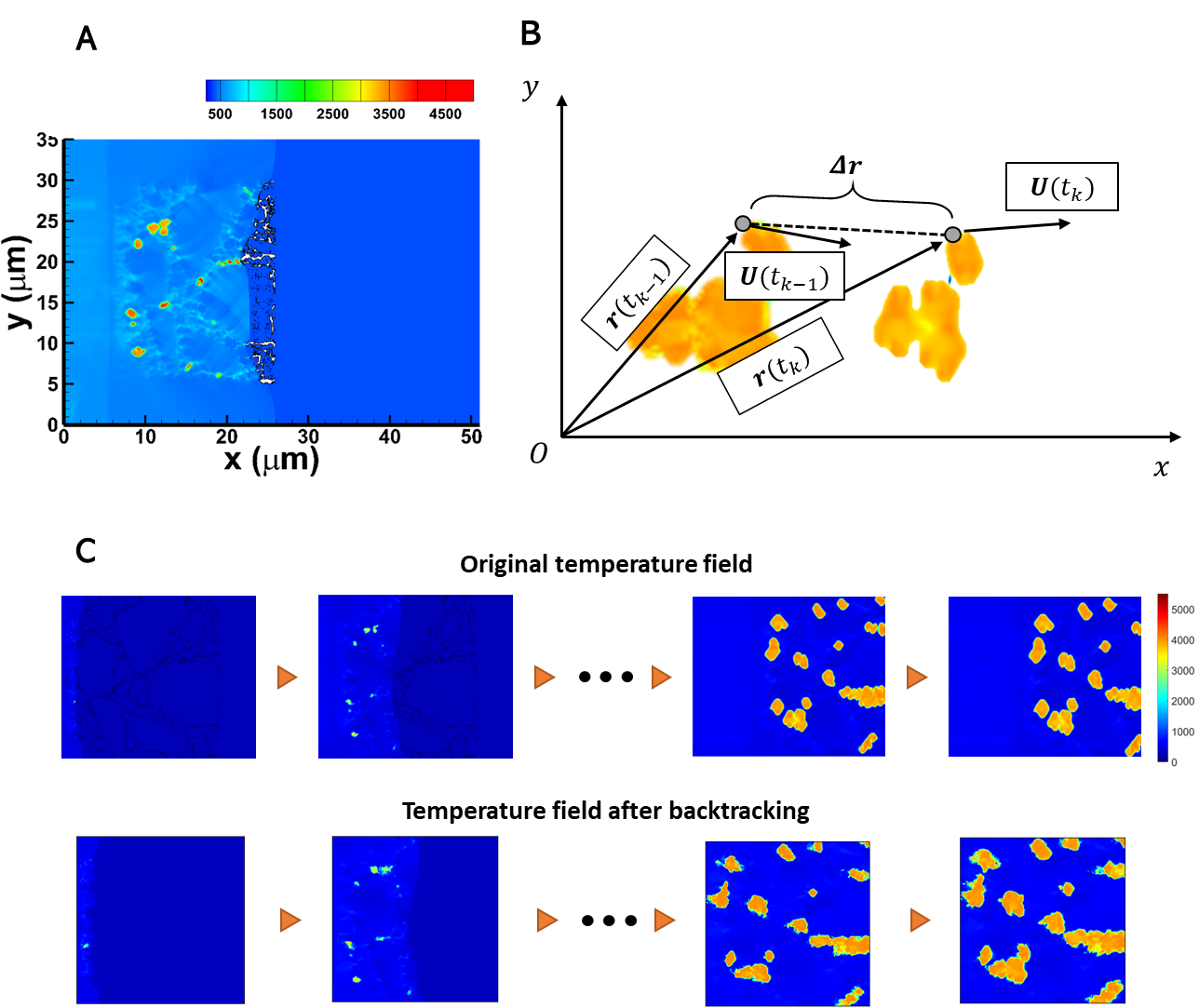}
    \caption{\textbf{Backtracking applied to temperature fields. (A)} The original DNS temperature fields include the deformation of microstructure, causing difficulty in PARC training. \textbf{(B)} The back-tracking algorithm using computed velocity field. \textbf{(C)} Comparison between original and temperature field after backtracking.}
    \label{fig:back_tracking}
\end{figure}

\subsection*{Post-processing of Simulation Data }
PARC uses a convolutional neural network (CNN) to associate the original microstructure (a field of voids) with the reactive response (hotspot ignition, growth, and coalescence). However, the original DNS-generated temperature and pressure fields also include the deformation of the microstructure caused by the application of shock load (Fig.~\ref{fig:back_tracking}A) which makes the training of PARC more difficult. As illustrated in Fig.~\ref{fig:back_tracking}A, as the shock passes through the domain, the voids are deformed, advected with the local particle velocity and collapse, eventually disappearing and being replaced by hotspots. The hotspots in turn are advected and deformed, assume complex shapes \cite{Sen2018a,Nguyen2022}, and grow to fuse with neighboring hotspots.  Such deformation effects in the state variables and the microstructure will cause difficulties for the prediction of PARC as the computational formulation of the model relies on regression on a fixed grid. For instance, if the hotspot is deformed to an extent greater than the size of CNN receptive field, it is nearly impossible for the CNN to predict the consequent QoI fields in this deformed setting. To overcome the potential difficulty caused by the deformation effect of the DNS-generated fields, we post-process simulation data to transform the original deformed DNS fields back into the non-deformed domain and microstructure. This post-processing of DNS data or ‘back-tracking’ (Fig.~\ref{fig:back_tracking}B) utilizes the DNS-computed velocity field associated with each snapshot of the simulations to calculate the microstructural deformation caused by applied shock. We recursively transform the QoI field in the deformed setting to the non-deformed one by calculating the displacement of every grid point in the domain from its position at the previous snapshot in the sequence of movie frames.

The backtracking of any field values from time step $t_k$ to a previous time step $t_{k-1}$ is conducted under the assumption that each material point lying at a grid node moves with (a known, DNS computed) velocity $\mathbf{u}$ associated with that grid node over the time interval $[t_{k-1},t_k ]$. The displacement $\Delta\mathbf{r}_k$ of a specific grid node associated with the material point at $t_k$ can be computed as
\begin{equation}
\label{eqn:delta_r}
  {\Delta}\mathbf{r}_k=\ 0.5\left(\mathbf{u}_{k-1}+\mathbf{u}_k\right)\Delta t
\end{equation}
where $\Delta t$ is the time interval between two video frames. Therefore, the position of the grid node, $\mathbf{r}_{k-1}$, at the previous time step $t_{k-1}$ can be computed from its position $\mathbf{r}_k$ at current time step $t_k$ as 
\begin{equation}
\label{eqn:backtracked}
  \mathbf{r}_{k-1}=\mathbf{r}_k-\Delta\mathbf{r}_k
\end{equation}

The computed current DNS-computed field is then collocated with the remapped ‘non-deformed’ one at that time step. Since the purpose of the backtracking process is to associate the DNS-computed field at a given time $t_k$ to the given initial microstructure at time $t_0$, the backtracking process is performed recursively from time $t_k$ to time $t_0$ as described below. Starting from the DNS-computed field at the discrete snapshot time $t_k$, follow the algorithm:
\begin{itemize}

	\item \textbf{Step 1}: Compute the displacement for every moving grid point of current time step k using Eq.~\ref{eqn:delta_r}.
	\item \textbf{Step 2}: Compute the position map of all moving grid points at the previous time step k-1 using Eq.~\ref{eqn:backtracked}
	\item \textbf{Step 3}: Transform the current QoI field to the undeformed configuration by assigning the current value to the remapped location provided in Eq.~\ref{eqn:backtracked} at the previous time step $k-1$
	\item \textbf{Step 4}: decrement time level counter $k$
	\item \textbf{Step 5}: If $k=0$, stop the process; if not, return to Step 1.
\end{itemize}

The result of the backtracking process is that the spatiotemporally evolving field of voids and hotspots is mapped back to the original fixed Cartesian grid so that the microstructure effectively evolves into a hotspot field in a domain that is undeformed (Fig.~\ref{fig:back_tracking}C).

\subsection*{Additional Results}
\begin{figure}[h]
    \centering
    \includegraphics[width=\textwidth]{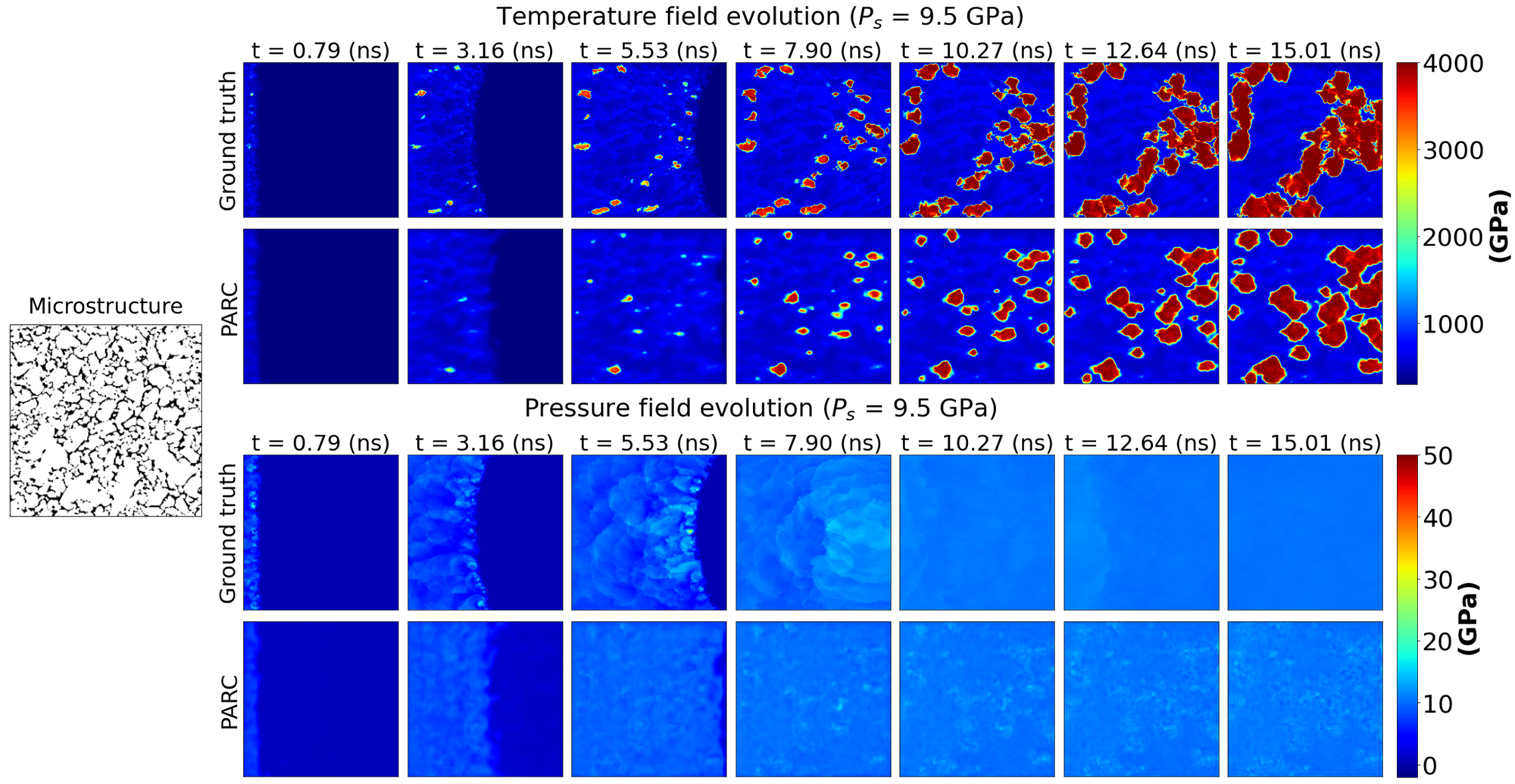}
    \caption{\textbf{Temperature and pressure field predictions for test microstructure \#2}}
    \label{fig:test_1}
\end{figure}

\begin{figure}[h]
    \centering
    \includegraphics[width=\textwidth]{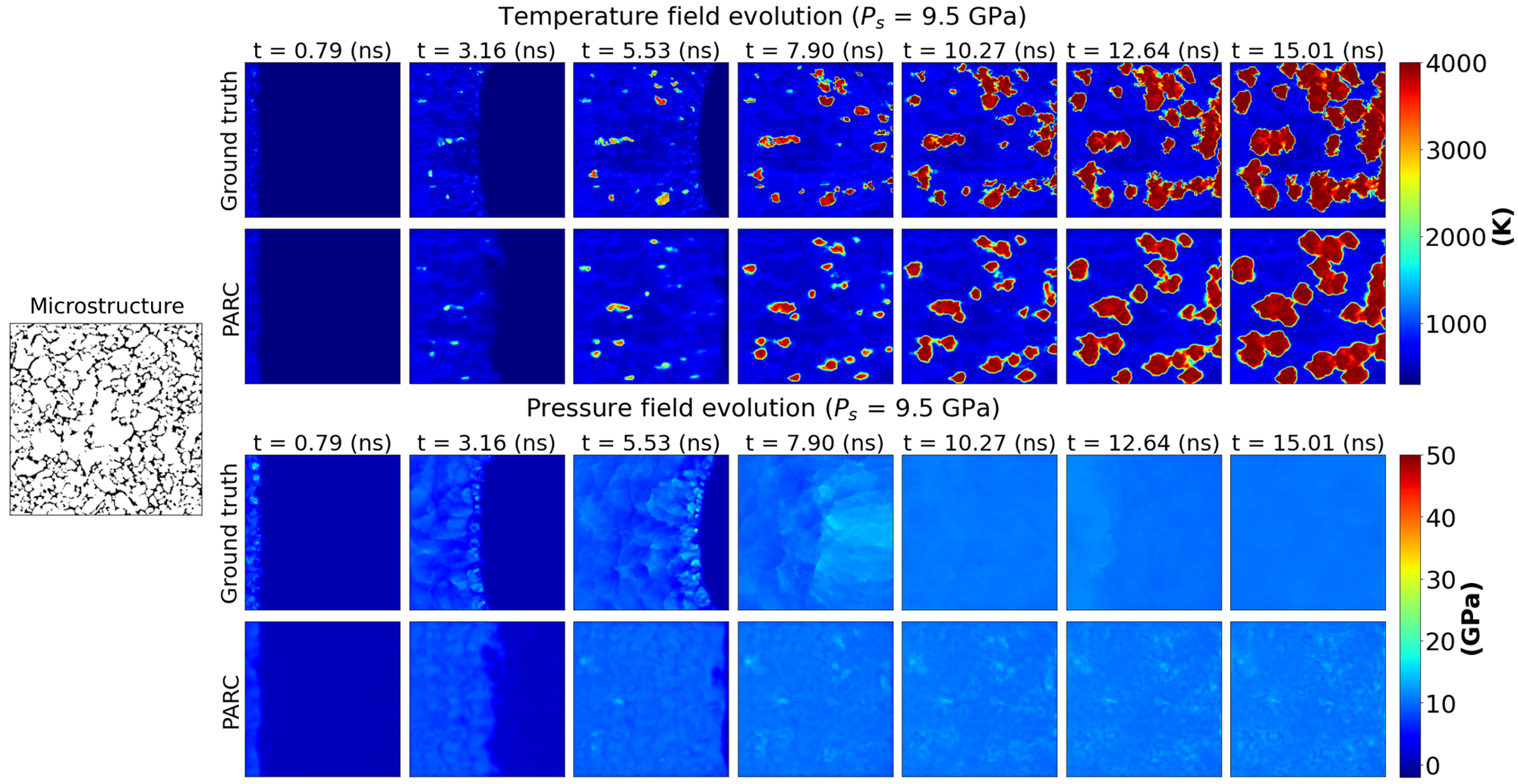}
    \caption{\textbf{Temperature and pressure field predictions for test microstructure \#3}}
    \label{fig:test_2}
\end{figure}

\begin{figure}[h]
    \centering
    \includegraphics[width=\textwidth]{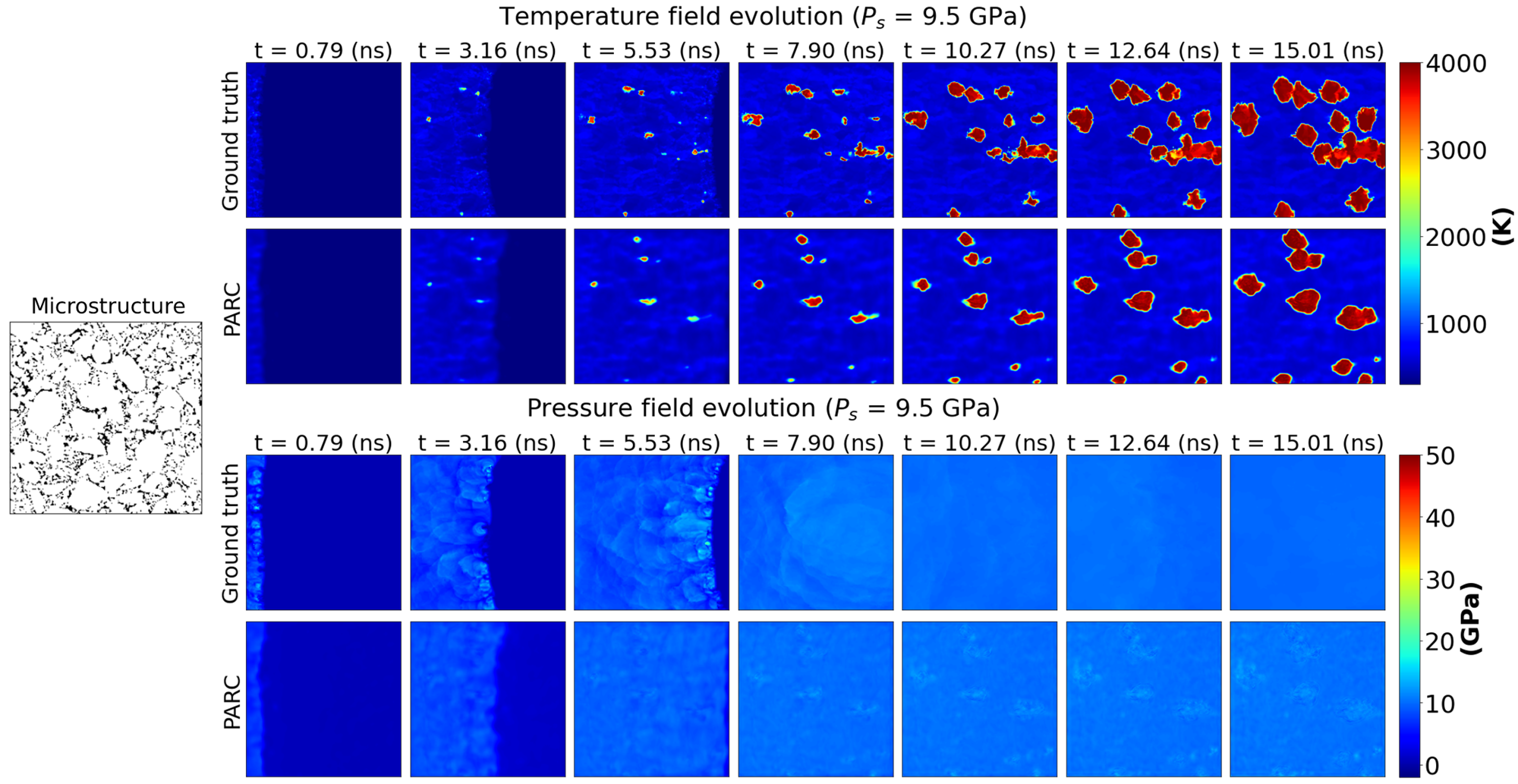}
    \caption{\textbf{Temperature and pressure field predictions for test microstructure \#4}}
    \label{fig:test_3}
\end{figure}

\begin{figure}[h]
    \centering
    \includegraphics[width=\textwidth]{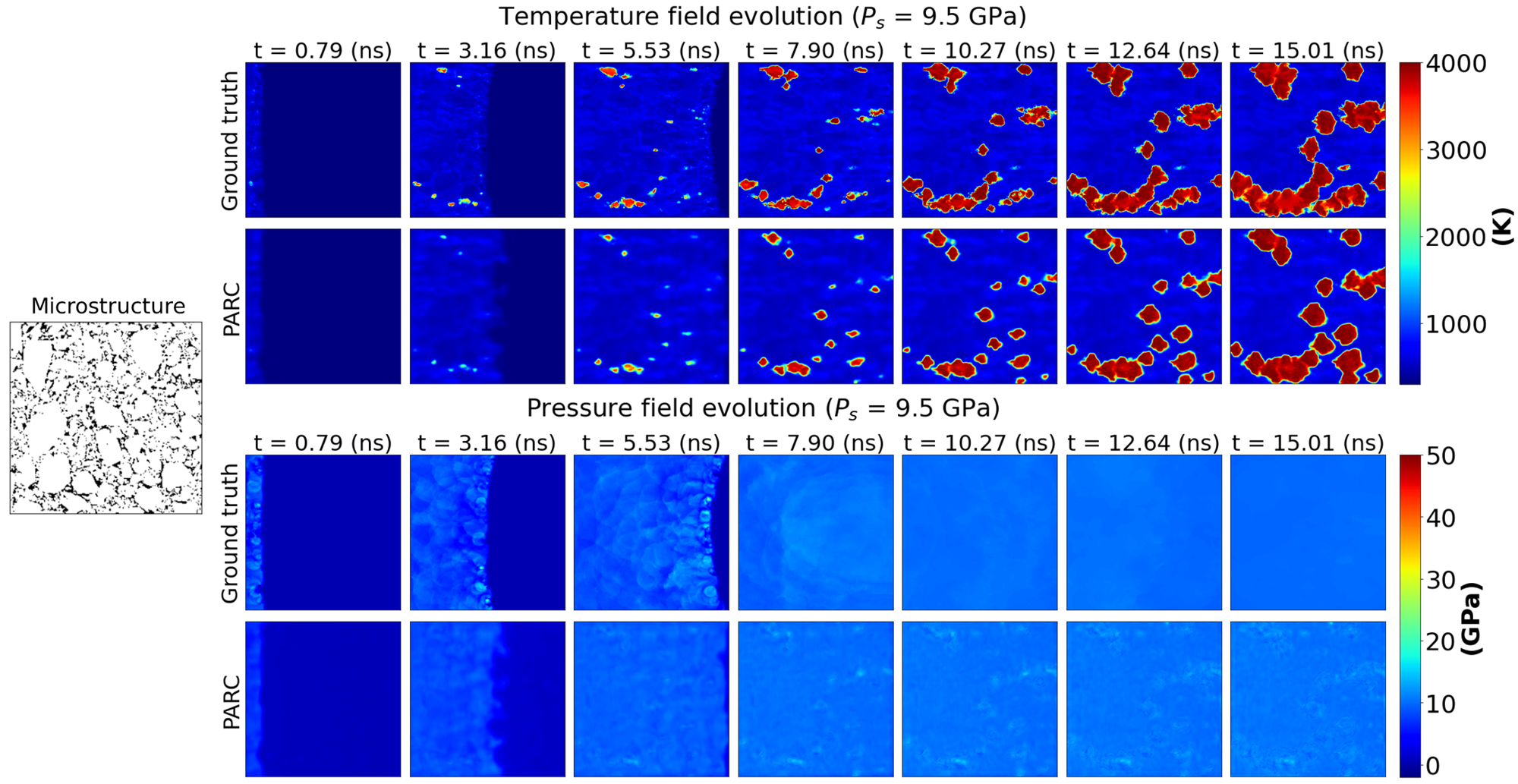}
    \caption{\textbf{Temperature and pressure field predictions for test microstructure \#5}}
    \label{fig:test_4}
\end{figure}

\begin{figure}[h]
    \centering
    \includegraphics[width=\textwidth]{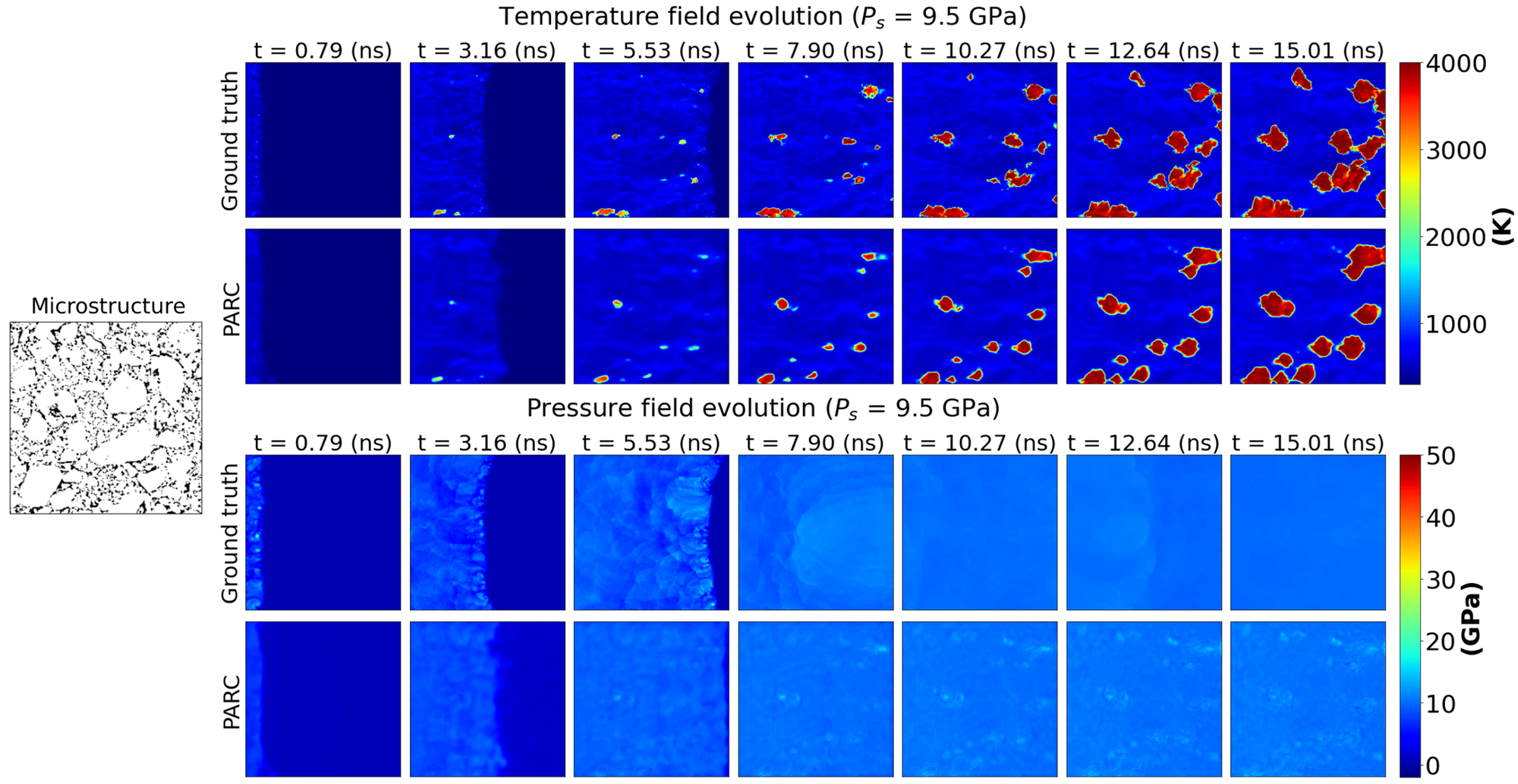}
    \caption{\textbf{Temperature and pressure field predictions for test microstructure \#6}}
    \label{fig:test_5}
\end{figure}

\begin{figure}[h]
    \centering
    \includegraphics[width=\textwidth]{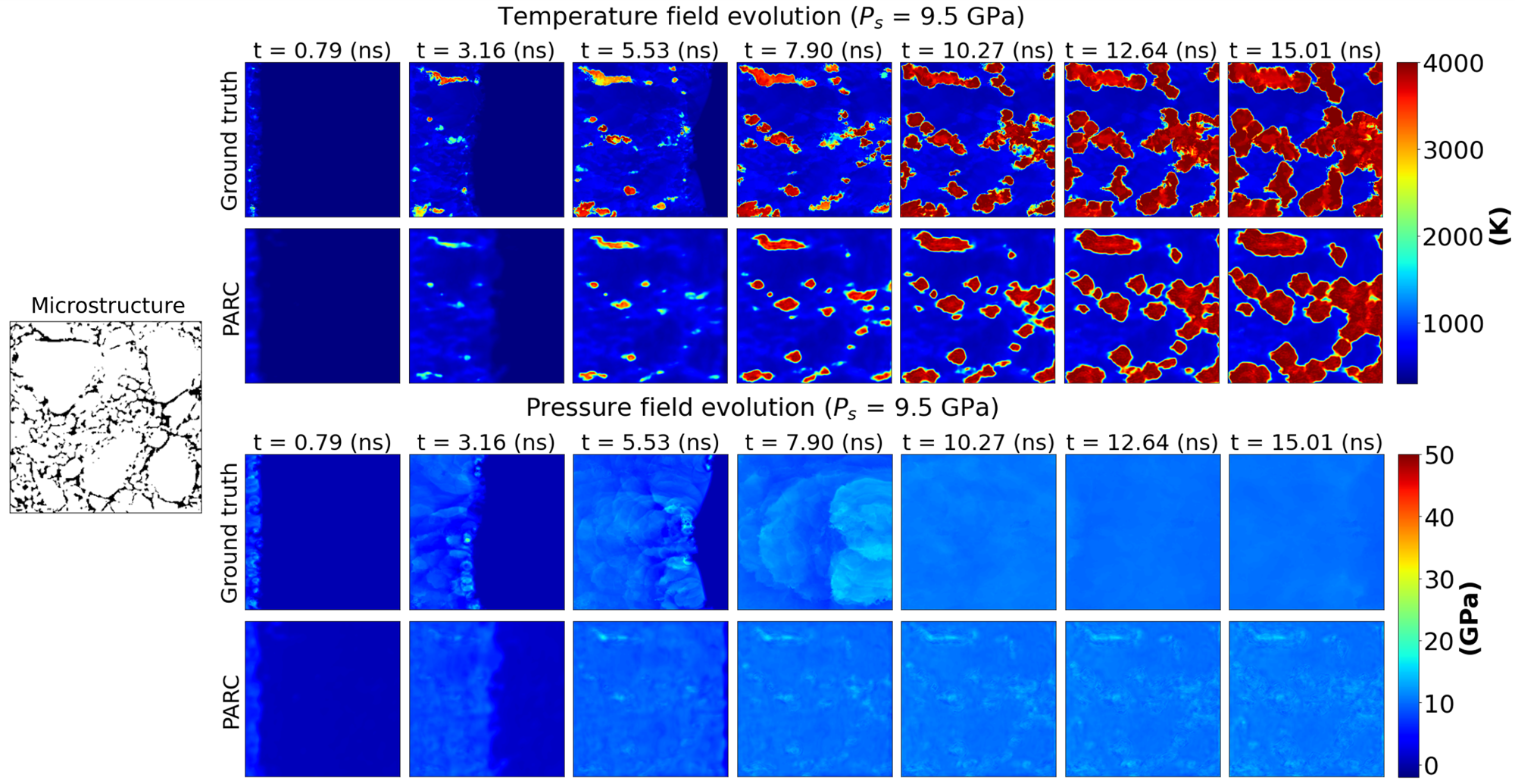}
    \caption{\textbf{Temperature and pressure field predictions for test microstructure \#7}}
    \label{fig:test_6}
\end{figure}

\begin{figure}[ht!]
    \centering
    \includegraphics[width=\textwidth]{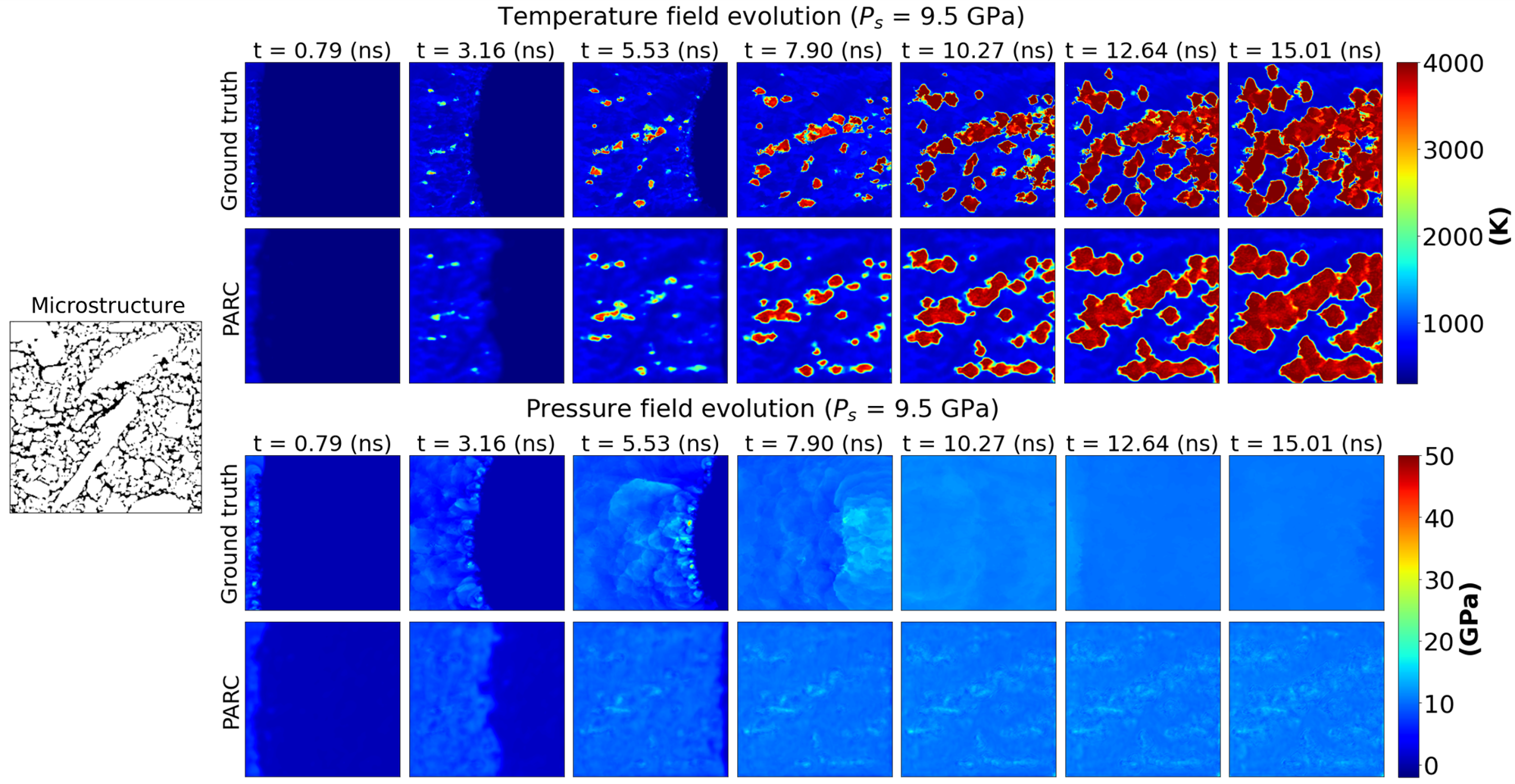}
    \caption{\textbf{Temperature and pressure field predictions for test microstructure \#8}}
    \label{fig:test_7}
\end{figure}

\begin{figure}[h]
    \centering
    \includegraphics[width=\textwidth]{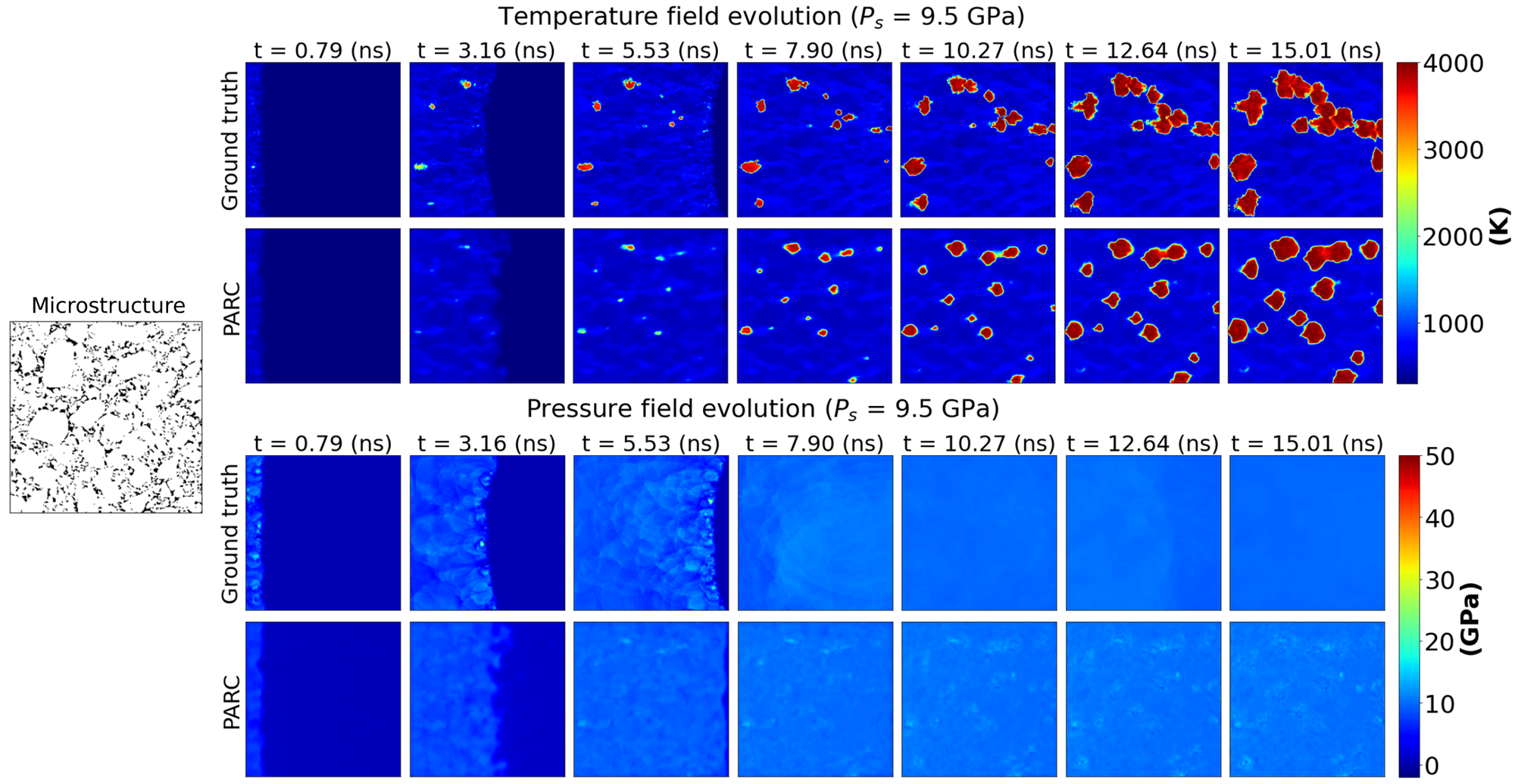}
    \caption{\textbf{Temperature and pressure field predictions for test microstructure \#9}}
    \label{fig:test_8}
\end{figure}

\begin{figure}[h]
    \centering
    \includegraphics[width=0.8\textwidth]{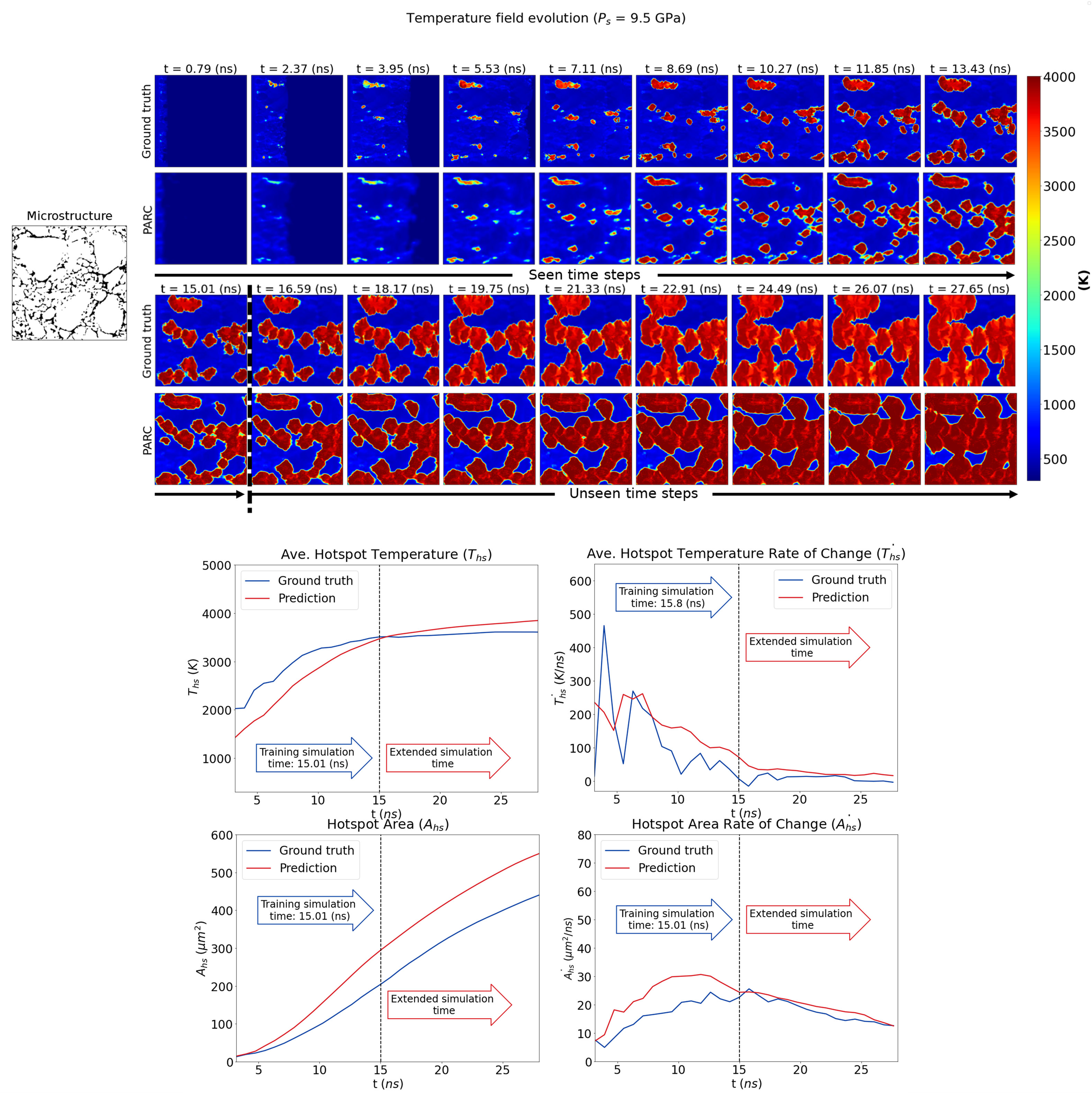} 
\caption{\textbf{PARC predictions in unseen time steps for EM microstructure sample \#2}}
\label{fig:PARC-result-extended-1}
\end{figure}

\begin{figure}[h]
    \centering
    \includegraphics[width=0.8\textwidth]{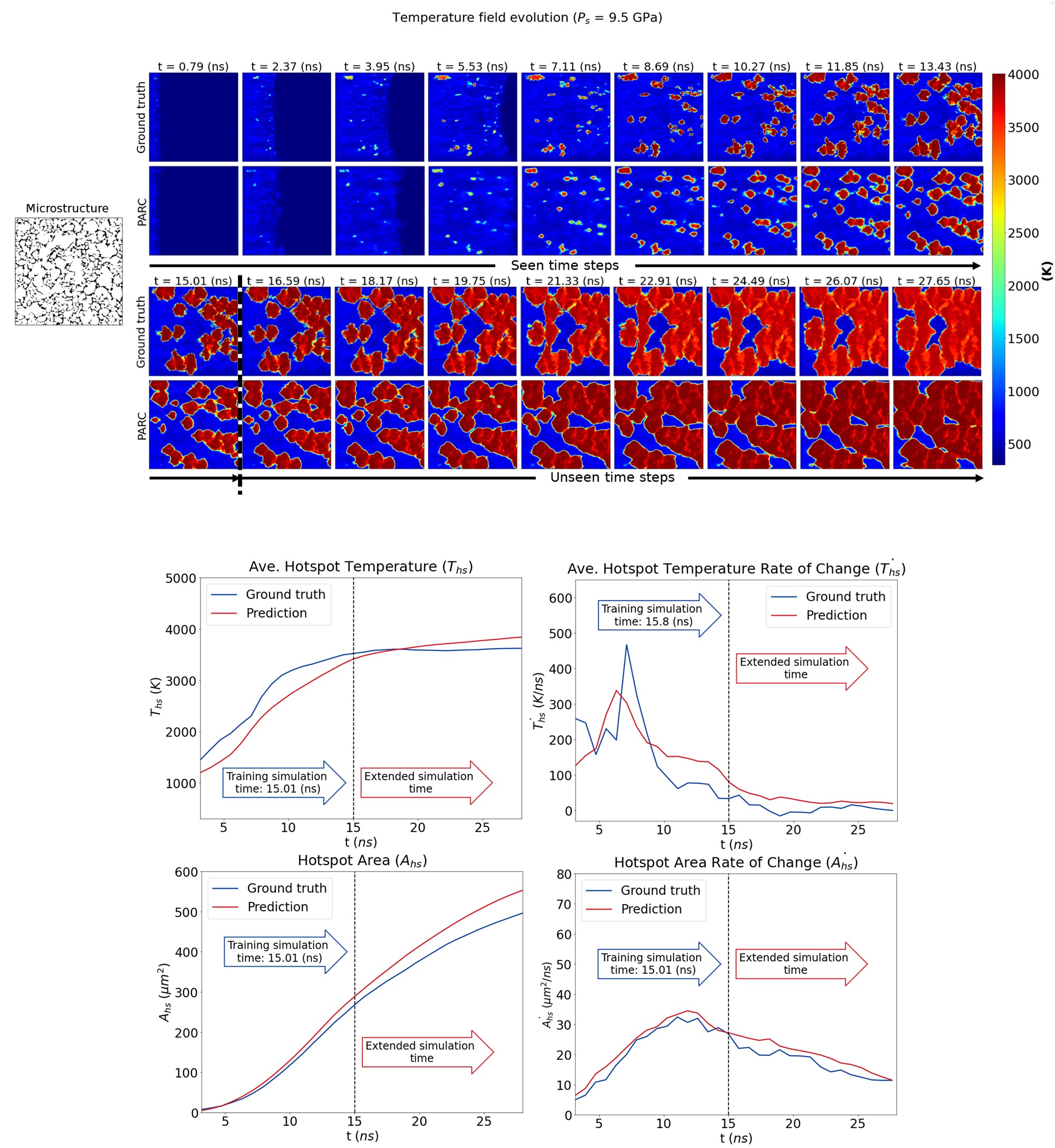} 
\caption{\textbf{PARC predictions in unseen time steps for EM microstructure sample \#3}}
\label{fig:PARC-result-extended-2}
\end{figure}

\begin{figure}[h]
    \centering
    \includegraphics[width=0.8\textwidth]{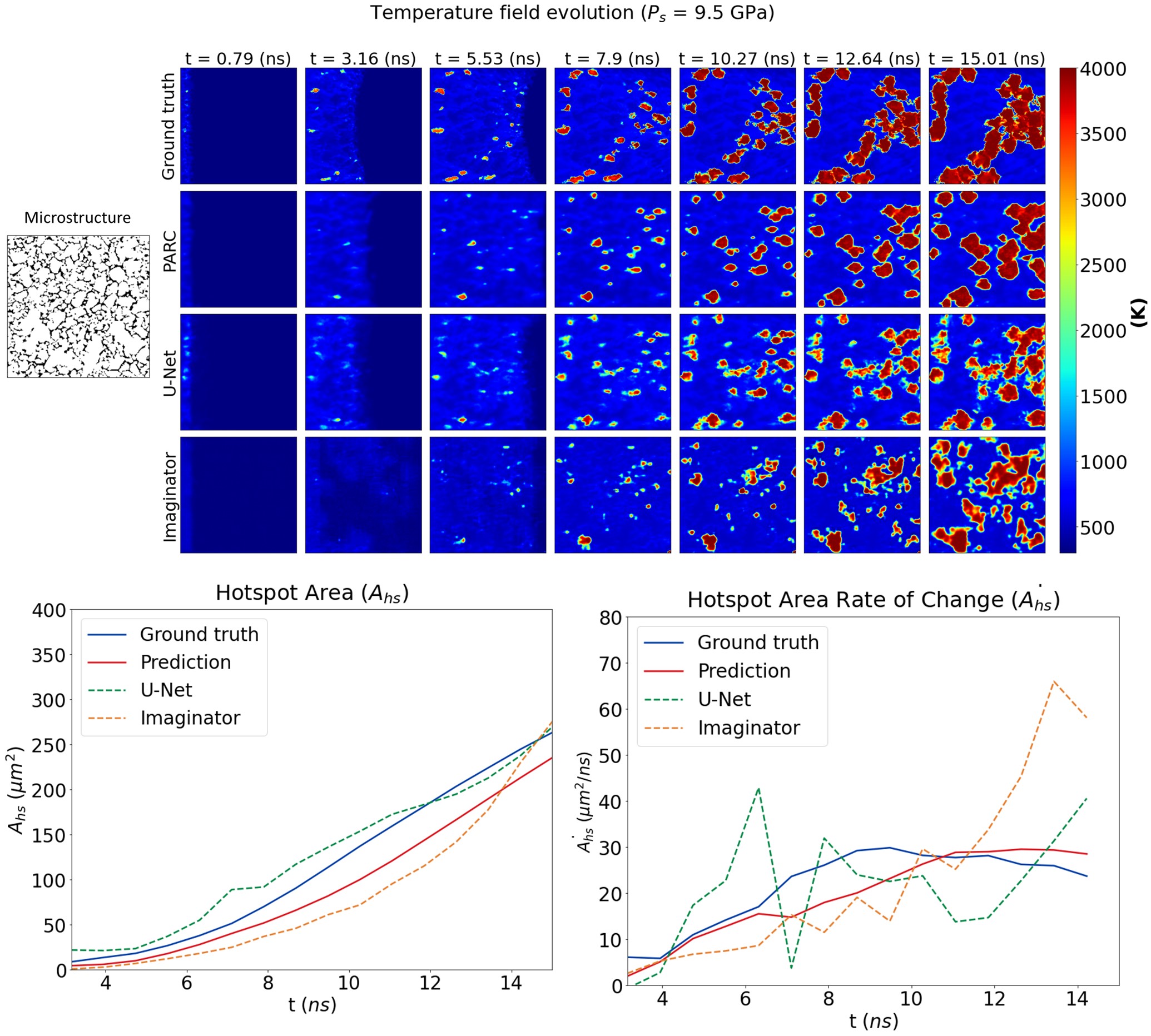} 
    \caption{\textbf{The benchmarking between PARC and other physics-na\"{i}ve ML models in hotspot evolution predictions for EM microstructure sample \#2}}
    \label{fig:four_model_compare_1}
\end{figure}

\begin{figure}[h]
    \centering
    \includegraphics[width=0.8\textwidth]{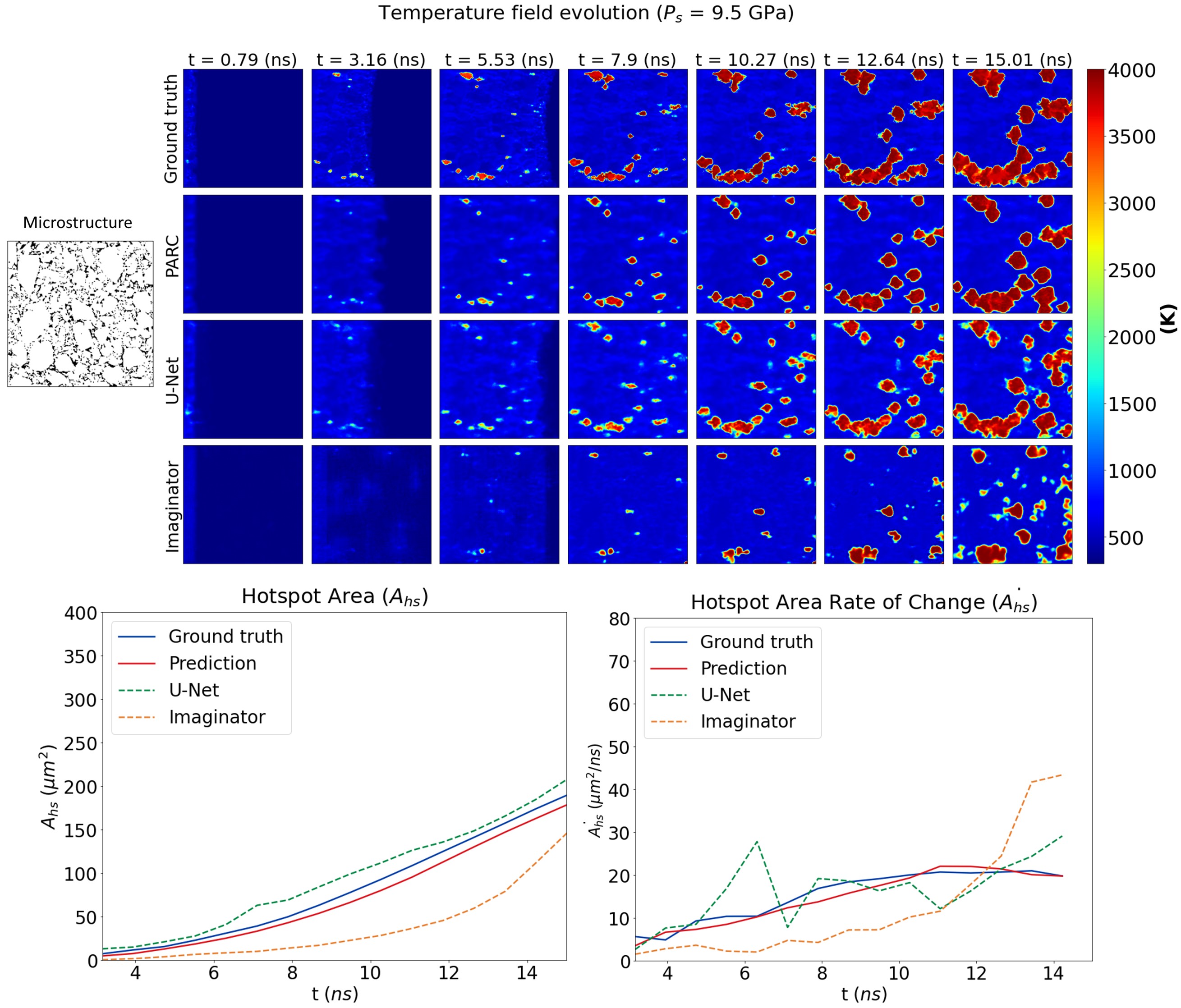} 
    \caption{\textbf{The benchmarking between PARC and other physics-na\"{i}ve ML models in hotspot evolution predictions for EM microstructure sample \#3}}
    \label{fig:four_model_compare_2}
\end{figure}

Additional results for PARC-predicted temperature and pressure field evolutions are given in Figs.~\cref{fig:test_1,fig:test_2,fig:test_3,fig:test_4,fig:test_5,fig:test_6,fig:test_7,fig:test_8}. Furthermore, additional results for the extended simulation time experiments are also given in Fig.~\cref{fig:PARC-result-extended-1,fig:PARC-result-extended-2}.
Finally, additional comparisons with the physics-na\"{i}ve benchmarks are given in Figs.~\cref{fig:four_model_compare_1,fig:four_model_compare_2}. 

\subsection*{Movie clips}
Besides the results presented in the figures, we also include movie clips describing the temperature and pressure field evolution predicted by PARC in comparison with ground truth DNS (Movies S.1 to S.9). Addtionally, temperature field evolutions before and after the backtracking procedure is given in Movie S.10.

\begin{enumerate}
    \item[\textbf{Movie S.1}] \textbf{Temperature and pressure field evolution movie for test microstructure $\#$1} 
    \item[\textbf{Movie S.2}] \textbf{Temperature and pressure field evolution movie for test microstructure $\#$2} 
    \item[\textbf{Movie S.3}] \textbf{Temperature and pressure field evolution movie for test microstructure $\#$3} 
    \item[\textbf{Movie S.4}] \textbf{Temperature and pressure field evolution movie for test microstructure $\#$4} 
    \item[\textbf{Movie S.5}] \textbf{Temperature and pressure field evolution movie for test microstructure $\#$5}
    \item[\textbf{Movie S.6}] \textbf{Temperature and pressure field evolution movie for test microstructure $\#$6} 
    \item[\textbf{Movie S.7}] \textbf{Temperature and pressure field evolution movie for test microstructure $\#$7} 
    \item[\textbf{Movie S.8}] \textbf{Temperature and pressure field evolution movie for test microstructure $\#$8} 
    \item[\textbf{Movie S.9}] \textbf{Temperature and pressure field evolution movie for test microstructure $\#$9} 
    \item[\textbf{Movie S.10}] \textbf{Temperature field before and after the backtracking procedure} 

\end{enumerate}

\end{document}